\documentclass[pra,showpacs,twocolumn,superscriptaddress,floatfix,aps]{revtex4-2}
\usepackage{amsmath,amssymb,bm}
\usepackage{graphicx}
\usepackage{physics}
\usepackage{color}
\usepackage{soul}
\usepackage{mathtools}
\usepackage[utf8]{inputenc}
\usepackage[T1]{fontenc}
\usepackage{array}
\usepackage{float}
\usepackage{comment}
\usepackage{appendix}
\begin{document}
\title{Ring Bose-Einstein condensate in a cavity: Chirality Detection and Rotation Sensing}
\author{Nalinikanta Pradhan}
\affiliation{Department of Physics, Indian Institute of Technology, Guwahati 781039, Assam, India}
\author{Pardeep Kumar}
\affiliation{Max Planck Institute for the Science of Light, Staudtstra\ss e 2, 91058 Erlangen, Germany}
\author{Rina Kanamoto}
\affiliation{Department of Physics, Meiji University, Kawasaki, Kanagawa 214-8571, Japan}
\author{Tarak Nath Dey}
\affiliation{Department of Physics, Indian Institute of Technology, Guwahati 781039, Assam, India} 
\author{M. Bhattacharya}
\affiliation{School of Physics and Astronomy, Rochester Institute of Technology, 84 Lomb Memorial Drive, Rochester, New York 14623, USA}
\author{Pankaj Kumar Mishra}
\affiliation{Department of Physics, Indian Institute of Technology, Guwahati 781039, Assam, India}

\date{\today}

\begin{abstract} 
Recently, a method has been proposed to detect the rotation of a ring Bose-Einstein condensate, \textit{in situ}, in real-time and with minimal destruction, using a cavity driven with optical fields carrying orbital angular momentum~\cite{KumarPRL2021,pradhan2023ringBEC}. This method is sensitive to the magnitude of the condensate winding number but not its sign. In the present work, we consider simulations of the rotation of the angular lattice formed by the optical fields and show that the resulting cavity transmission spectra are sensitive to the sign of the condensate winding number. We demonstrate the minimally destructive technique on persistent current rotational eigenstates, counter-rotating superpositions, and a soliton singly or in collision with a second soliton. Conversely, we also investigate the sensitivity of the ring condensate, given knowledge of its winding number, to the rotation of the optical lattice. This characterizes the effectiveness of the optomechanical configuration as a laboratory rotation sensor. Our results are important to studies of rotating ring condensates used in atomtronics, superfluid hydrodynamics, simulation of topological defects and cosmological theories, interferometry using matter-wave solitons, and optomechanical sensing.

\end{abstract}

\flushbottom

\maketitle
\section{Introduction}
Degenerate atoms contained in a ring-shaped potential are paradigms of quantum rotation \cite{RyuPRL2007,GuoPRL2020,CaiPRL2022,DelPacePRX2022,GuoNJP2022}. Specifically, a Bose-Einstein Condensate (BEC) in a ring displays quantized persistent superflow \cite{MoulderPRA2012}, phase slips \cite{WrightPRL2013,SnizhkoPRA2016}, solitons \cite{KanamotoPRA2003}, hysteresis \cite{EckelNature2014,HouPRA2017}, excitations \cite{WrightPRA2013}, spinor hydrodynamics \cite{BeattiePRL2013,GallemiNJP2015},  and shock waves \cite{WangNJP2015}. Such a BEC can be used to simulate topological systems \cite{KanamotoPRL2008,DasScientificReports2012,CormanPRL2014}, black holes \cite{YatsutaPRR2020}, the early universe \cite{EckelPRX2018}, and time crystals \cite{OhbergPRL2019}. Apart from fundamental interest,
ring BEC's are platforms critical to applications such as matter-wave interferometry \cite{GuptaPRL2005,MartiPRA2015}, gyroscopy \cite{CooperPRA2010,PelegriNJP2018}, atomtronics \cite{RamanathanPRL2011,RyuPRL2013,PandeyPRL2021,AmicoRMP2022}, quantum computation \cite{AghmalyanNJP2015}, and rotation sensing \cite{RagolePRL2016}. 

In all of the investigations mentioned above, a central role is played by the state of rotation of the condensate. It is no surprise that therefore a number of methods exist for probing the condensate winding number, or topological charge, which is the angular momentum per atom divided by Planck's constant $\hbar$. In the earliest instances, the method involved measuring the diameter of the central density hole of the atomic distribution upon time-of-flight expansion
and subsequent absorption imaging \cite{MurrayPRA2013, BeattiePRL2013}. This diameter is proportional to the magnitude of the initial condensate vortex charge. However, the method does not yield the sign of the winding number, which is the direction of the atomic rotation. Another technique, which provides the same information, involves interferometric processing of the condensate so that the image contains a number of density peaks equal in number to the winding number \cite{MoulderPRA2012}.

Subsequently, methods that revealed the sign of the winding number were demonstrated. One way of implementing such detection is to use a disk-shaped reference condensate, which is also allowed (along with the ring) to expand via time-of-flight and the matter-wave phase interferogram resulting from its interference with the ring carries a clear signature of the handedness of the atomic rotation. Specifically, if the ring is not rotating, the interferogram fringes consist of a set of concentric circles. If the ring is rotating, the fringes appear in the shape of nested spirals 
whose number and handedness reveal the magnitude and sign, respectively, of the condensate winding number \cite{CormanPRL2014}. 

However, all demonstrated methods of detecting ring BEC rotation are fully destructive of the condensate because they all employ absorption imaging \cite{KumarNJP2016}. The absorption and subsequent emission of photons destroy the coherence of the atomic condensate. In recent work, our group proposed a method for detecting atomic rotation with minimal destruction to the condensate in real-time and \textit{in situ} \cite{KumarPRL2021}. Our proposal involved coupling the rotating ring BEC to a resonator using optical beams carrying orbital angular momentum (OAM) [see Fig. 1] and subsequently using the well-established detection techniques of cavity optomechanics \cite{BrenneckeScience2008,AspelmeyerRMP2014, AbbottPRL2016}. While the initial proposal modeled the BEC as a two-mode system, in subsequent work we used a mean-field model that accounts for all the modes of the BEC and considered solitons as well as persistent currents \cite{pradhan2023ringBEC}.  

Nevertheless, the minimally destructive technique proposed by us only reveals the magnitude, and not the winding number, of the ring BEC. This may be understood from a symmetry point of view by realizing that the optical beams introduced by us produce an angular lattice overlapping with the ring BEC (see Fig. 1). Since this lattice breaks rotational symmetry about the cavity axis, it can probe the angular momentum, i.e., the winding number, of the condensate. However, since it does not break the chiral symmetry by picking a handedness about the cavity axis, it is insensitive to the sign of the condensate rotation.

In the present work, we show that the rotation of the optical lattice, which can be carried out in the laboratory using spatial light modulators \cite{HeOESLMOL}, allows us to determine the sign of the condensate rotation. For realistic lattice rotation frequencies, we demonstrate our technique on persistent current rotational eigenstates and counter-rotating superpositions for weak and repulsive atomic interactions. We also show how the method works for solitons, singly or in colliding pairs, for weak and attractive atomic interactions \cite{BrandJPB2001,KavoulakisPRA2003,ToikkaJPB2013,McDonaldPRL2014,HelmPRL2015,GalluciNJP2016,
JezekPRA2016,CataldoEPJD2016}. Finally, we also consider the converse of the problem, where we think of the lattice rotation as caused by the laboratory and ask how well a condensate with a known winding number can measure this rotation. In other words, we consider the sensitivity of the ring BEC in a cavity as a rotation sensor. 

We present analytical few-mode calculations, where possible, and numerical simulations that 
treat the condensate in the mean-field limit and the cavity field classically. Damping and noise arising from the matter as well as optical fields have been modeled realistically. Our results include cavity transmission spectra, which establish rotation (sign as well as magnitude) measurement, the sensitivity of the measurement as a function of system response frequency, and atomic density profiles showing the effect of the measurement on the condensate.
Before concluding, we mention that a theoretical proposal for measuring the magnitude as well as the sign of the condensate winding number using atom counting exists \cite{SafaeiPRA2019}. 

This paper is organized as follows: In Section~\ref{TheorModelNumSim} the theoretical model and details of the numerical simulation are presented. In Sections \ref{sec: PersistentCurrent} and \ref{sec: Soliton} we provide the dynamics, OAM content, optical spectra, and measurement sensitivity for persistent current and bright soliton detection, respectively. In Sec.~\ref{sec: CritO} we present a detailed analysis of the critical rotation required for using the ring BEC as a rotation sensor. Following this, an analysis of the fidelity is presented in Sec.~\ref{sec:Fidelity}. Finally, we conclude our work in Sec.~\ref{sec:5}. In the Appendix~\ref{App:highRot}, we present the chirality detection and rotation sensing at high lattice rotation frequency for both persistent current and soliton state of BEC.

\section{Theoretical model and details of numerical simulation}
\label{TheorModelNumSim}
In this section, we describe the few-mode quantum mechanical model for the configuration of interest and the mean-field equations, which take all modes of the condensate into account.
\subsection{Few-mode Hamiltonian}
\label{FewModeHam}
\begin{figure*}[!htp]
\begin{center}
    \includegraphics[width=0.6\linewidth]{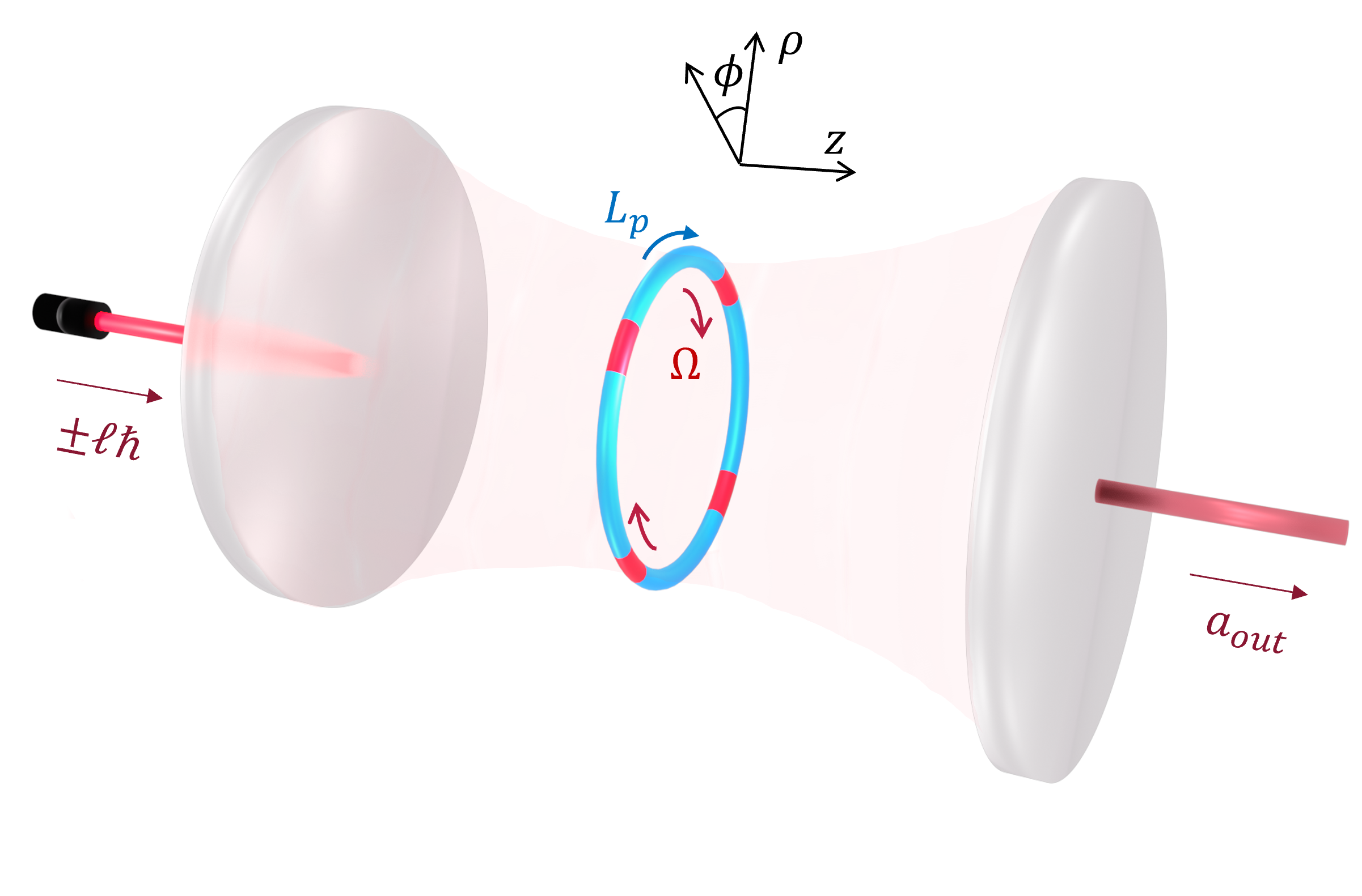} \\

\end{center}
\caption{Schematic illustration of a Bose-Einstein condensate with winding number $L_p$ trapped in a ring. Two Laguerre-Gauss cavity modes carrying orbital angular momenta $\pm \ell \hbar $ form an annular lattice to probe the dynamics of the condensate. The lattice is rotated at frequency $\Omega$. The optical field transmitted by the cavity is denoted by $a_{out}$.}
\label{fig:setup}
\end{figure*}

We consider a BEC confined in a ring trap of radius $R$, whose center lies on the axis of an optical resonator; see Fig.~\ref{fig:setup}. The BEC is probed by a superposition of frequency-degenerate Laguerre-Gaussian beams carrying optical OAM $\pm \ell\hbar$. The combination of these beams creates an angular lattice inside the cavity, overlapping with the ring BEC. The matter waves in the condensate Bragg diffract from this lattice, giving rise to persistent current sidemodes $L_{p}\rightarrow L_{p}\pm 2\ell$, where $L_{p}$ is the winding number of the supercurrent initially present in the BEC. 

In the rotating frame of the laser drive and the rest frame of the optical lattice, the azimuthal dynamics of the BEC are described in terms of the Hamiltonian written in second quantized form
\begin{align}
    \hat{H}_0 &= \int_0^{2\pi}  \Psi^{\dagger} (\phi) \left[\left(-i\frac{d}{d\phi}\right)^2 + U_0 \cos^2 (\ell\phi) a^{\dagger}a\right]\Psi(\phi)d\phi\nonumber\\
    & +\frac{{\cal G}}{2}\int_0^{2\pi} \Psi^{\dagger}(\phi)\Psi^{\dagger}(\phi)\Psi(\phi)\Psi(\phi)d\phi \nonumber\\
    &- \Delta_o a^{\dagger}a -i\eta (a-a^{\dagger})\;,\label{Eq:RestHamil}
\end{align}
where $\Psi(\phi)$ is the bosonic atomic field operator such that $[\Psi(\phi),\Psi^{\dagger}(\phi^{\prime})]=\delta(\phi-\phi^{\prime})$, and $\phi$ is the angular variable of atomic position along the ring. The optical field operators, on the other hand, obey $[a,a^{\dagger}]=1$. The square bracket in the first line of Eq. (\ref{Eq:RestHamil}) has two terms: (i) the first term represents the rotational kinetic energy of the atoms, and (ii) the second contribution governs the interaction of the atoms with the optical lattice potential such that $U_0=g_0^{2}/\Delta_a$, where $g_0$ and $\Delta_{a}$ are the single photon-atom coupling strength and detuning of the optical drive from the atomic transition, respectively. The two-body atomic interaction with the strength $\mathcal{G}=g/\hbar\omega_{\beta}$ is represented in the second line of Eq.~(\ref{Eq:RestHamil}) which corresponds to binary collisions in the condensate. Here  $g=2\hbar\omega_{\rho}a_{s}/R$ depends on the atomic $s$-wave scattering length $a_s$ and the harmonic trap frequency along the radial direction $\omega_{\rho}$; further, $\omega_{\beta}$ is defined to scale the energy such that $\hbar\omega_{\beta}=\hbar^{2}/(2mR^{2})$, where $m$ is the atomic mass. The terms in the last line of Eq.~(\ref{Eq:RestHamil}) contain contributions from the cavity field energy and the cavity drive, respectively. Here $\Delta_{o}$ is the drive detuning from the cavity resonance $\omega_{o}$ and $\eta=\sqrt{P_{in}\gamma_o/\hbar\omega_o}$ is the drive strength, where $P_{in}$ ($\gamma_o$) is the input optical power (cavity linewidth).

\subsection{Rotating Angular Lattice}
As the angular lattice breaks the rotational symmetry of the system about the cavity axis, it can be used to probe the 
winding number $L_{p}$ of the ring condensate. However, the lattice does not break the chiral symmetry of the system by providing a preferred handedness, and only the magnitude of $L_{p}$ can be found using this method, as shown earlier \cite{KumarPRL2021}. 

To detect the \textit{sign} of the persistent current, we break the chiral symmetry of the system by rotating the optical lattice with an angular frequency of $\Omega$. Mathematically, this corresponds to replacing 
$\cos^{2}\left(\ell\phi\right)$ on the first line of Eq.~(\ref{Eq:RestHamil}) by $\cos^{2}\left(\ell\phi+\Omega t\right).$ Using then the operator $e^{i\Omega\hat{L}_{\phi}t}$ that transforms the Hamiltonian from the laboratory to the rotating frame of the optical angular lattice, where  
\begin{align}
   \hat{L}_{\phi} &= \int_0^{2\pi}  \Psi^{\dagger} (\phi) \left(-i\frac{d}{d \phi}\right) \Psi(\phi)d\phi\;.\label{Eq:AngMoment}
\end{align}
is the angular momentum operator, we obtain   
\begin{align}
    \hat{H}(\Omega)&=\hat{H}_{0}-\Omega\hat{L}_{\phi}\;.\label{Eq:RotHamil}
\end{align}
Using Eqs.~(\ref{Eq:RestHamil}) and (\ref{Eq:AngMoment}) in Eq.~(\ref{Eq:RotHamil}) and neglecting the constant terms, the Hamiltonian in the rotating optical lattice frame becomes \cite{KanamotoPRA2003}
\begin{align}
    \hat{H}(\Omega) &= \int_0^{2\pi}  \Psi^{\dagger} (\phi) \left[\left(-i\frac{d}{d\phi}-\frac{\Omega}{2}\right)^2\right]\Psi(\phi)d\phi\nonumber\\ 
    & + \int_0^{2\pi}  \Psi^{\dagger} (\phi) U_0 \cos^2 (\ell\phi) a^{\dagger}a\Psi(\phi)d\phi\nonumber\\
    & +\frac{{\cal G}}{2}\int_0^{2\pi} \Psi^{\dagger}(\phi)\Psi^{\dagger}(\phi)\Psi(\phi)\Psi(\phi)d\phi \nonumber\\
    &- \Delta_0 a^{\dagger}a -i\eta (a-a^{\dagger})\;. \label{Eq:FinalRotHamil}
\end{align}
Physically, the atoms of the condensate get Bragg diffracted from the optical lattice. We will consider photon numbers smaller than unity in the cavity; hence, the lattice is weak, and only first-order atomic diffraction is non-negligible. This results in two sidemodes, $\omega_{c}$ and $\omega_{d}$, respectively, in terms of the matter-wave OAM states. The frequencies of these sidemodes can be deduced by following the procedure outlined in \cite{KumarPRL2021}, with a straightforward modification arising from the transformation between laboratory and lattice rotation frames 
\begin{align}
    \omega_{c,d}(\Omega)&=\omega_{\beta}\left(L_p\pm 2\ell-\frac{\Omega'}{2}\right)^{2}\;,\label{Eq:Freq}
\end{align}
where the normalized angular frequency is $\Omega' = \Omega / \omega_\beta$. From Eq.~(\ref{Eq:Freq}) it follows that for a fixed angular frequency $\Omega'$ of lattice rotation, the frequencies of the sidemodes are different for opposite winding numbers $\pm L_{p}$. Thus, this technique can be used to infer the direction of the BEC rotation. In contrast, in the absence of lattice rotation $(\Omega'=0)$, it can be verified that transforming $L_{p}\leftrightarrow -L_{p}$ merely exchanges the sidemode frequencies ($\omega_{c} \leftrightarrow \omega_{d})$, thus leaving the sign undetermined. In the above discussion, we have neglected the effect of atomic interactions on the sidemode frequencies. Inclusion of interactions leads to the modified frequencies \cite{KanamotoPRA2003,KumarPRL2021}
\begin{equation}
\label{eq:OmP}
\omega_{c,d}'=\left[\omega_{c,d}\left(\omega_{c,d}+4\tilde{g}N\right)\right]^{1/2},    
\end{equation}
where $\tilde{g}=g/(4\pi\hbar)$ and $N$ is the number of atoms in the condensate.
We will be comparing the analytical expression of the sidemode frequencies from Eqs.~(\ref{eq:OmP}) to our numerical simulations presented below. 

We note that the optical lattice rotates through the condensate, presenting obstacles that are penetrable by the superfluid, since the peak optical potential $U_{0}|\alpha_{s}|^{2}$ (where $\left|\alpha_s\right|^2$ is the steady state photon number in the cavity) is always smaller than the chemical potential $\mu$. In this work, we will only consider lattice rotation rates smaller than the speed of sound $v_{s}=\sqrt{\mu/m}$ in the condensate, i.e., such that $\omega_{\beta}\Omega' < v_{s}/(2\pi R)$. This ensures that the flow around the ring is always superfluidic, and the local Landau criterion for the onset of dissipative flow is never satisfied \cite{Lim2022Shedding}.

\subsection{Equations of Motion}
In contrast to the few-mode model discussed above, we now take into account the full-mode characterization of the condensate by employing a mean-field Gross-Pitaevskii formalism. Using Eqs.~(\ref{Eq:FinalRotHamil}), the classical mean-field equation can be derived, including the fluctuations \cite{stoof1999coherent,stoof2001dynamics}. Then the coupled dynamical equations for the condensate wave function $\psi$ and light field amplitude $\alpha$ in the rotating optical lattice are written as~\cite{pradhan2023ringBEC}

\begin{equation}
\begin{aligned}
    (i-\Gamma)\frac{d\psi}{d\tau} = &\biggl[-  \frac{d^2}{d\phi^2} + \frac{U_0}{\omega_\beta}  |\alpha(\tau)|^2 {\cos^2\left({\ell\phi}\right)} -\mu   \\ & + \mathcal{G}\,|\psi|^2 -\Omega' (-i \frac{d}{d \phi}) \biggr] \psi + \xi(\phi,\tau),
    \label{Eq:bec}    
\end{aligned}
\end{equation}
and
\begin{equation}
\begin{aligned}
    i\frac{d\alpha}{d\tau} = \biggl\{ - \biggl[\Delta_c &- U_0 \langle {\cos^2\left({\ell\phi}\right)}\rangle_{\tau} + i \frac{\gamma_{0}}{2}\biggr] \alpha  + i\eta \biggr\} \omega_\beta^{-1}\\ & + i \sqrt{\gamma_0} \omega_\beta^{-1} \alpha_{in}(\tau). 
    \label{Eq:cavity}
\end{aligned}
\end{equation}
Eqs.~(\ref{Eq:bec}) is the dimensionless stochastic Gross-Pitaevski equation, where $\psi \equiv \psi(\phi, \tau)$ represents the condensate wave function, which is normalized to the total number of atoms in the condensate $N$, as expressed by
\begin{equation}
    \int_{0}^{2 \pi} |\psi(\phi,\tau)|^2  d\phi = N.
    \label{Eq:scale_wf}
\end{equation}
Here $\phi$ and $\tau$ are the angular coordinates along the ring and the scaled time respectively. The length is normalized to the radius of the ring $R$, while the energy and time are normalized as $\hbar \omega_{\beta} = \hbar^2 / 2 m R ^2$ and $\tau = \omega_\beta t\;$ respectively. Since most of the terms used in Eqs. (\ref{Eq:bec}) and (\ref{Eq:cavity}) have been described in earlier sections, only a brief discussion of the terms associated with the fluctuations is presented here. The optical fluctuation is taken into account through the term $\alpha_{in}(\tau)$ and the thermal fluctuation associated with the condensate is through $\xi(\phi,\tau)$, which is related to the damping of the condensate  $\Gamma$ (scaled with $\omega_\beta$) according to the fluctuation-dissipation theory \cite{kubo1966fluctuation} and to conserve the norm of the condensate in presence of this fluctuation, the chemical potential $\mu$ is corrected at each time step as~\cite{mithun2018signatures}
\begin{equation*}
    \Delta\mu = (\Delta \tau)^{-1} \ln{\left[\int \lvert \psi(\phi, \tau)\rvert^2 d\phi / \int \vert\psi(\phi, \tau + \Delta \tau)\rvert^2 d\phi\right]}.    
\end{equation*}
Here both the thermal and optical noise are considered to be delta-correlated white noise, with the correlations~\cite{das2012winding,KumarPRL2021,pradhan2023ringBEC}
\begin{align} 
\langle\xi(\phi,\tau) \, \xi ^ *(\phi',\tau')\rangle  &= \frac{2\,\Gamma \,k_B \,T}{\hbar \omega_\beta} \, \delta(\phi - \phi') \, \delta (\tau - \tau'), \\ 
\langle \alpha_{in}(\tau) \, \alpha_{in}^ *(\tau') \rangle  &= \omega_\beta \, \delta (\tau - \tau').
\end{align}
So the terms $\xi(\phi,\tau)$ and $\alpha_{in}(\tau)$ can be modelled as
\begin{align} 
\xi(\phi,\tau)  &= \sqrt{\frac{2\,\Gamma \,k_B \,T}{\hbar \omega_\beta d \phi \, d \tau}} \, \mathcal{N}(0,1,N_{\phi}) \, \mathcal{N}(0,1,N_{\phi}), \\ 
\alpha_{in}(\tau)  &= \sqrt{\omega _\beta / d\tau} \,\mathcal{N}(0,1,1), 
\label{eq:noise_alpha}
\end{align}
where $k_B$ is the Boltzmann constant and $T$ is the temperature of the bath. Here $\mathcal{N}(0,1,N_{\phi})$ and $\mathcal{N}(0,1,1)$ are the sets of random variables that are normally distributed, having zero mean and unit variance. The third argument in $\mathcal{N}$ denotes the size of the array, containing the random numbers that are added in each time step with the pre-factor while solving the above-mentioned coupled differential equations. 



\subsection{Simulation details}
In this work, we have considered four different ground states of the condensate, namely, persistent current, superposition state, soliton, and two soliton states moving with equal and opposite angular velocity. First, in Section~\ref{sec: PersistentCurrent}, we demonstrate the dynamics of a single persistent current and two counter-propagating persistent currents in a super-positioned state. Subsequently, in Section~\ref{sec: Soliton}, we present the dynamics of a bright soliton and the dynamics of a pair of solitons that undergo multiple collisions. The dynamics of the persistent current are captured by solving the coupled set of dynamical equations (Eqs. (\ref{Eq:bec}) and (\ref{Eq:cavity})) numerically using the real-time dynamics scheme. We use the Fourier pseudo-spectral method~\cite{dutykh2016brief} aided by the fourth-order Runge-Kutta scheme for the temporal evolution of the condensate~\cite{tan2012general}. For persistent current, we start the simulation by considering the initial state as $e^{i L_p \phi}$. However, to generate a soliton-like ground state, we first evolve the condensate with the Gaussian state $e^{-\phi^2 / 2}$, which resembles the shape of a bright soliton, using the imaginary time scheme based upon the Strang splitting Fourier method \cite{bao2003numerical}. Subsequently, the soliton initial state is modulated with a phase $e^{i L_p \phi}$ to achieve the moving soliton state, which is further evolved using the real-time scheme.

For all the simulation runs, the dynamics of condensate and cavity are captured for a period of $5$ sec that gives a single trajectory of the cavity field, which is later used to obtain the cavity output spectrum through a Fourier transformation. For enhanced visualization, the cavity output spectrum is smoothed using the moving average technique of a window size of 30 Hz. The rotation measurement sensitivities have been calculated by fitting the output spectrum near the frequency at which we get the peaks with the appropriate shape. This process allows us to reduce the effect of background noise coming from the frequency away from the relevant ones in the calculation of the sensitivity. For all the simulation runs, we have chosen the $dt$ as $10^{-7}$ where the space resolution $d\phi$ is set at $0.006$.

\section{Results and discussion}

\subsection{Persistent current}
\label{sec: PersistentCurrent}

\subsubsection{Rotational eigenstate}
We consider a condensate comprised of $N$ number of $^{23}$Na atoms \cite{RyuPRL2007} each of mass $m$, confined in an annular trap generating a persistent current. The macroscopic condensate wave function, representing the rotational state of persistent current, can be assumed to have the form of a plane wave, having the initial form as 
\begin{equation}
\label{eq:psi1}
    \psi(\phi) = \sqrt{\frac{N}{2 \pi}} e^{i L_p \phi}.
\end{equation}
Here $e^{i L_p \phi}$ is the phase factor, which presents a phase gradient to the condensate of uniform density, and $L_p$ represents the winding number of the condensate, which is the quantity to be detected.

\begin{figure}[b]
\begin{center}
    \includegraphics[width=1\linewidth]{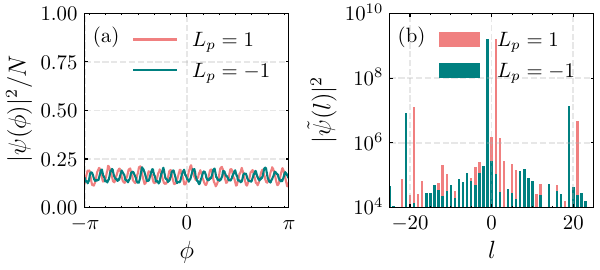} \\
\end{center}

\caption{Persistent current: (a) angular profile of the condensate density per particle for a persistent current rotational eigenstate; (b) OAM state content of the condensate. Parameters used here are $\Omega' = 0.5 $, $ L_p = \pm 1$, $\Gamma = 0.0001$, $ T = 10$ nK, $g / \hbar \simeq 2 \pi \times 0.02 $ Hz, $\ell = 10$, $N = 10^4$, $\Tilde{\Delta} = -2 \pi \times 173$ Hz, $ U_0 = 2\pi \times 212$ Hz, $ \gamma_{0} = 2\pi\times 2$ MHz, $P_{in} = 50$ fW, $\Omega' = 0.5$, $\omega_c = 2\pi \times 10^{15}$ Hz, $m = 23$ amu, and $R = 12$ $\mu$m. }
  \label{fig:PW1}
\end{figure}

As our present method can detect the magnitude as well as the sign of the winding number of the persistent current, in this section, we present the simulation results for two different currents having winding number $\pm L_p$ for specific values of the angular frequency of the rotating optical lattice $\Omega'$. In Fig.~\ref{fig:PW1} we show the ground state density profiles obtained for $L_p=\pm 1$ with the rotation frequency $\Omega'=0.5 $, along with the occupation of matter wave OAM states for both values of $L_p$.

We find that the condensate density profiles are quite similar to each other and appear to be slightly modulated as a result of the rotating optical lattice and the thermal noise present in the system [See Fig.~\ref{fig:PW1} (a)]. To illustrate the presence of different angular momentum modes we show the Fourier transformation of the density of the persistent currents in Fourier space corresponding to $L_p = \pm 1$ in Fig.~\ref{fig:PW1} (b). Here the side modes at the position $L_p \pm 2l - \Omega'/2$ appear due to matter wave diffraction.

\begin{figure}[b]
\centering
\includegraphics[width=1\linewidth]{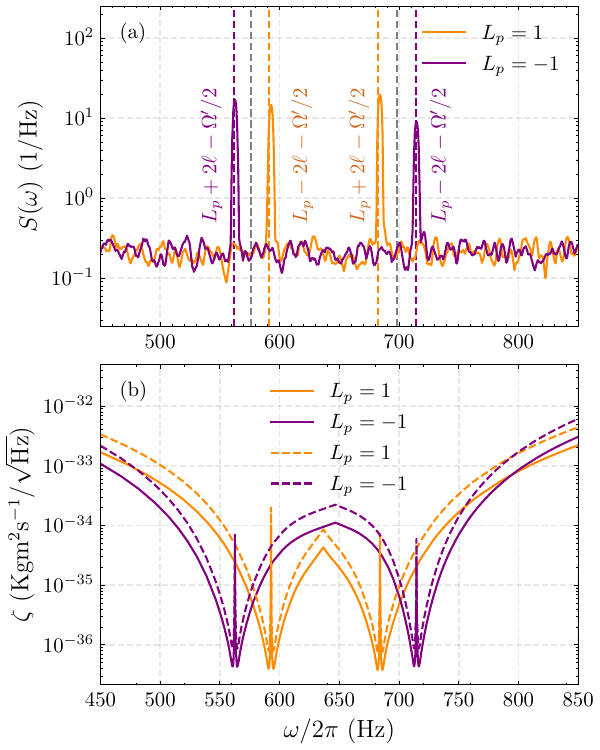} \\
\caption{Persistent current: (a) Noise spectra of the output phase quadrature of the cavity field and (b) Rotation measurement sensitivity versus the system response frequency for different winding numbers $L_{p}=\pm1$. In (a), the gray-colored vertical dashed line corresponds to the analytical predictions for the side modes for $\Omega' = 0$ (provided as a reference to indicate the opposite direction of shifts for $L_{p}=\pm 1$) and the orange and purple-colored vertical dashed lines correspond to the analytical predictions for the side modes for $\Omega' = 0.5$. In (b), the solid curves correspond to $\zeta^\Lambda$ (Eq.~\ref{eq:Sensitivity_Lp1}), while the dashed curves correspond to $\zeta^\Omega$ (Eq.~\ref{eq:Sensitivity_Lp2}). The other parameters used here are the same as mentioned in Fig.~\ref{fig:PW1}.}
  \label{fig:PW2}
\end{figure}

To probe the different OAM components present in the condensate, we compute the spectrum of the output optical field, which is transmitted through the cavity. Within the cavity, the light field is modulated at the side mode frequencies of the condensate $\omega_{c,d}'(\Omega)$, Eq.~(\ref{eq:OmP}) \cite{BrenneckeScience2008, KumarPRL2021}. To reveal these side mode frequencies, we perform a Fourier transform of the cavity field amplitude of the output field, which is related to the intra-cavity field through the input-output relation of cavity optomechanics as $\alpha_{out} = -\alpha_{in} + \sqrt{\gamma_0}\alpha$ \cite{AspelmeyerRMP2014}. For all the analysis performed in the paper, we consider the phase quadrature of the cavity transmission spectrum defined as
\begin{equation}
\label{eq:Spect}
S(\omega)=\left|\mathrm{Im}\left[\alpha_{out}(\omega)\right]\right|^{2}.
\end{equation}

Fig.~\ref{fig:PW2} (a) depicts the phase quadrature of the cavity transmission spectrum as a function of the response frequency of the system for two different situations (the same as in Fig.~\ref{fig:PW1}), i.e., for the winding numbers $L_p=\pm 1$. The spectrum clearly shows two distinct peaks at the locations corresponding to the side mode frequencies $\omega_{c,d}'(\Omega)$ for $L_p = \pm 1$. 
The numerically simulated peak positions match quite well with the analytical predictions of Eq.~(\ref{eq:OmP}).

\begin{figure}[b]
\centering
\includegraphics[width=1\linewidth]{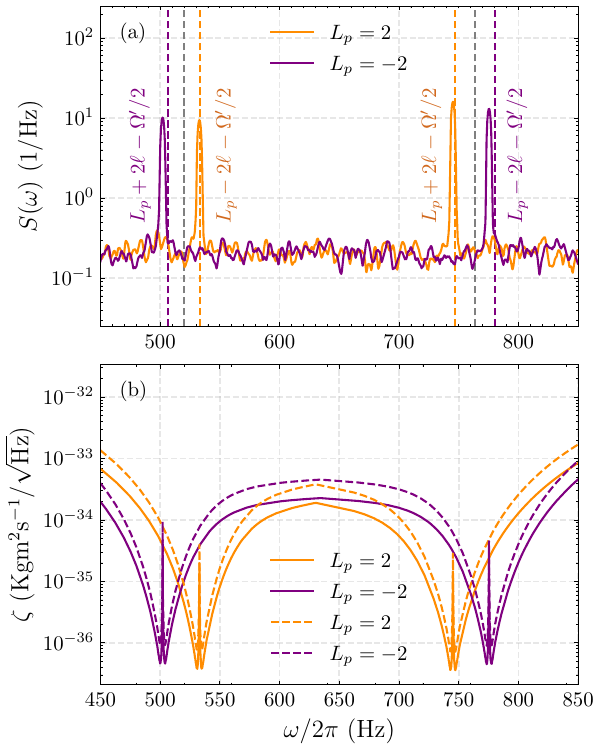} \\
\caption{Persistent current: (a) Noise spectra of the output phase quadrature of the cavity field, (b) Rotation measurement sensitivity as a function of the system response frequency for $L_p = \pm 2$. In (a), the gray-colored vertical dashed line corresponds to the analytical predictions for the side modes for $\Omega' = 0$ and the orange and purple-colored vertical dashed line corresponds to the analytical predictions for the side modes for $\Omega' = 0.5$. In (b), the solid curves correspond to $\zeta^\Lambda$ (Eq.~\ref{eq:Sensitivity_Lp1}), while the dashed curves correspond to $\zeta^\Omega$ (Eq.~\ref{eq:Sensitivity_Lp2}). The other set of parameters used here are the same as mentioned in Fig.~\ref{fig:PW1}.}
  \label{fig:PW3}
\end{figure}

Next, we compute the sensitivity of detecting $L_p$ for a fixed value of angular frequency of rotating optical lattice $\Omega'$, defined as

\begin{equation}
\label{eq:Sensitivity_Lp1}
    \zeta^\Lambda = \frac{S(\omega)}{\partial S(\omega)/\partial\Lambda} \times \sqrt{t_{meas}},
\end{equation}
and also the sensitivity of measuring the rotation of the optical lattice $\Omega$ for a fixed value of $L_p$ defined as
\begin{equation}
\label{eq:Sensitivity_Lp2}
    \zeta^\Omega = \frac{S(\omega)}{\partial S(\omega)/\partial (\hbar \Omega')} \times \sqrt{t_{meas}}.
\end{equation}
Here $t_{meas}^{-1} = 8 (\alpha_s G)^2 / \gamma_0$ is the optomechanical measurement rate, $ G = U_0 \sqrt{N} / 2 \sqrt{2}$ \cite{KumarPRL2021}, and $\Lambda = \hbar L_p $ is the angular momentum of the condensate. In Fig.~\ref{fig:PW2} (b), we show the sensitivities $\zeta^\Omega$ and $\zeta^\Lambda$, respectively, as functions of the system rotation frequency. The sensitivities of detecting the condensate winding number $L_p$ and rotation of the optical lattice $\Omega$ are optimized near the side mode frequencies $\omega_{c,d}'(\Omega)$.

To demonstrate that our method is effective at detecting two neighboring winding numbers, we further compute the cavity output spectrum for $L_p = \pm 2$ as shown in Fig.~\ref{fig:PW3} (a), where the peaks are spectrally distinct from the peaks for the case of $L_p = \pm 1$ [Fig.~\ref{fig:PW2}(a)]. The corresponding sensitivities of measurement are illustrated in Fig.~\ref{fig:PW3} (b).

After discussing the cavity spectra and sensitivities for the persistent current, in what follows, we present a detailed behavior of these quantities for the situation when we have the presence of the superposition of two persistent current states.

\subsubsection{Two state superposition}

\begin{figure}[b]
\includegraphics[width= 1 \linewidth]{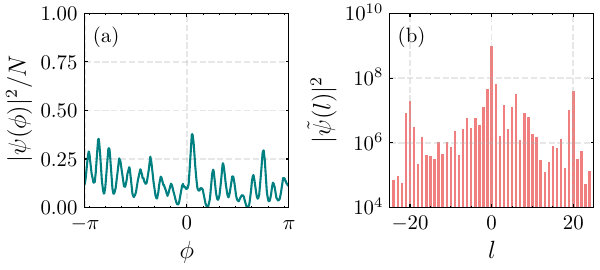} \\
\caption{Persistent current superposition: (a) Condensate density profile per particle with $L_{p1} = 1, L_{p2} = -1$, (b) OAM distribution of the condensate. Here $P_{in} = 0.7$ pW and the other parameters used are the same as in Fig.~\ref{fig:PW1}.}
\label{fig:PW5}
\end{figure}

\begin{figure*}[t]

\includegraphics[width= 0.9 \linewidth]{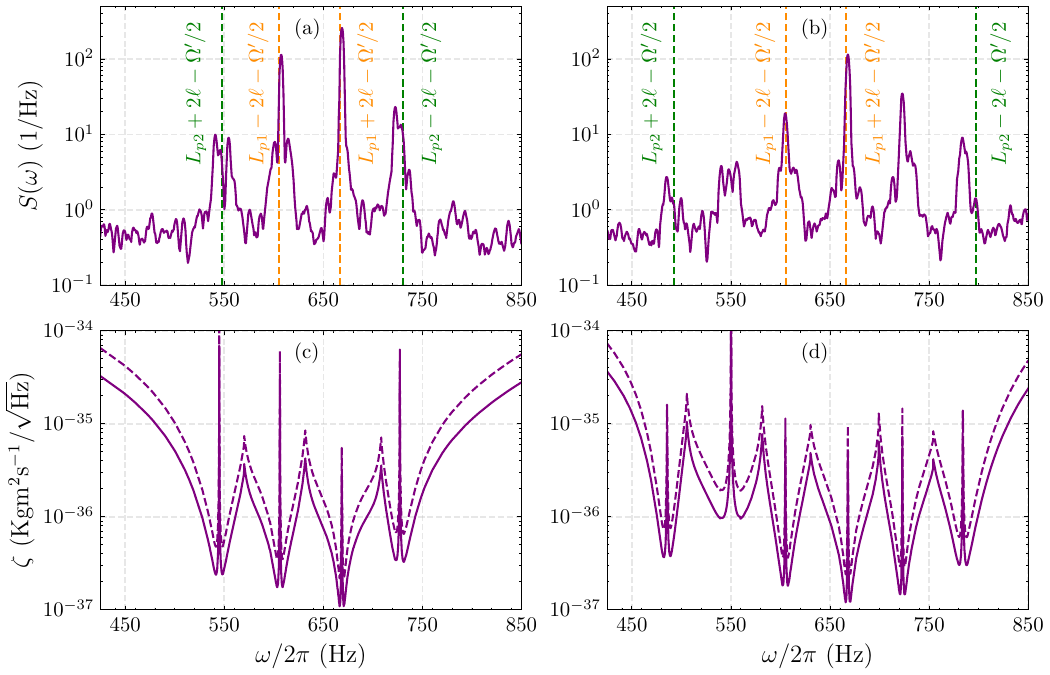} \\
\caption{Persistent current superposition: Left column for $L_{p1}=1$ and $L_{p2}=-1$  and right column for $L_{p1}=1$ and $L_{p2}=-2$. (a) and (b) Noise spectrum of the output phase quadrature versus response frequency, (c) and (d) Rotation measurement sensitivity versus response frequency. The solid curves correspond to $\zeta^\Lambda$ (Eq.~\ref{eq:Sensitivity_Lp1}), while the dashed curves correspond to $\zeta^\Omega$ (Eq.~\ref{eq:Sensitivity_Lp2}). The vertical dashed lines correspond to analytical predictions for $L_{p1}\pm 2l- \Omega'/2$ and $L_{p2}\pm 2l - \Omega'/2$. The set of parameters used is the same as in Fig.~\ref{fig:PW5}.}
\label{fig:PW9}
\end{figure*}

As our method allows us to detect both the magnitude and sign of the winding number associated with the persistent current, in this section we illustrate the capability of detecting winding numbers of two counter-propagating persistent currents for two different scenarios: one having equal magnitude but opposite signs, and the other involving different magnitudes and opposite signs. For these cases, we consider the initial state as 

\begin{equation}
\label{eq:Super}
   \psi(\phi) = \sqrt{\frac{N}{4 \pi}} 
   \left(e^{i L_{p1}\phi}+e^{iL_{p2}\phi}\right),
\end{equation}
which is a superposition state of two plane waves having winding numbers $L_{p1}$ and $L_{p2}$, respectively.

Figure~\ref{fig:PW5} shows the condensate density profile and matter-wave OAM distribution of the condensate wave function, representing the superposition of two counter-propagating persistent currents having winding numbers $L_{p1}=1$ and $L_{p2}=-1$. The increased modulation in condensate density is expected due to the superposition involving constructive and destructive interference of the two components of the superposition. The OAM distribution shows the dominant states that correspond to $L_p = 0$ and $\pm 2 \ell$, respectively, and these peaks serve as evidence for the interference between $L_{p1}=1$ and $L_{p2}=-1$ currents. Due to this interference, the occupation of other modes is relatively high compared to Fig.~\ref{fig:PW1} (b).

Figure~\ref{fig:PW9} (a) and (b) show the noise spectra of the output phase quadrature as a function of response frequency for $L_{p1}=1$, $L_{p2}=-1$ and $L_{p1}=1$, $L_{p2}=-2$, respectively. The peaks detected in the output spectrum represent the side mode frequencies [$\omega_{c,d}'(\Omega)$] of the two persistent currents in the superposition state. Additionally, some extra peaks are visible, which result from the interference between these two currents. The values of the winding numbers of the constituent persistent currents were determined uniquely by locating the dominant peaks (which yield $L_{p1}$) and the outermost peaks (which yield $L_{p2}$). In implementing this procedure, careful selection of input optical power plays a pivotal role, since it not only determines the visibility of peaks in the output spectrum but also regulates the operation of the system in the linear response regime. The increased noise in the cavity output spectra of Figs.~\ref{fig:PW9} (a) and (b), as compared to the single persistent current cases [Figs.~\ref{fig:PW2}(a) and ~\ref{fig:PW3}(a)] is the consequence of employing higher input optical power. A detailed discussion on the effect of higher input optical power on the cavity output spectrum can be found in \cite{bhattacharya2015rotational,pradhan2023ringBEC}.


\subsection{Soliton}
\label{sec: Soliton}
\subsubsection{Single Soliton}

A soliton refers to a self-bound localized state that propagates in a medium without any dispersion, and this localization is a result of the attractive interaction among the atoms constituting the soliton~\cite{Khawaja2002,Gangwar2022}. In the present work, we study the dynamics of a matter-wave soliton forming due to the condensation of $N$ number of $^7$Li atoms having a negative scattering length, leading to an attractive inter-atomic interaction. In particular, we detect the rotation of this soliton in the ring trap, where the optical lattice is rotating with an angular frequency $\omega$. This allows us to measure the winding number associated with the soliton rotation.
\begin{figure}[b]
\centering
\includegraphics[width=1\linewidth]{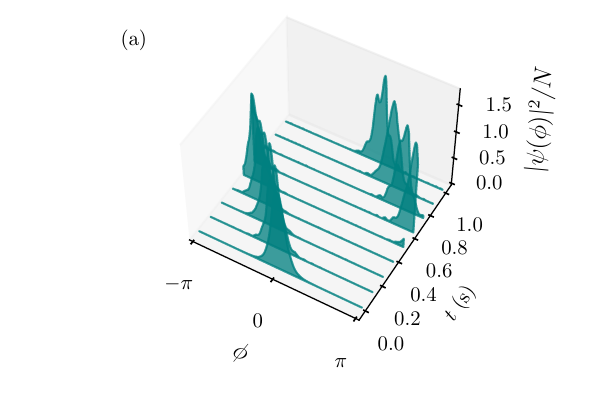} \\
\includegraphics[width=1\linewidth]{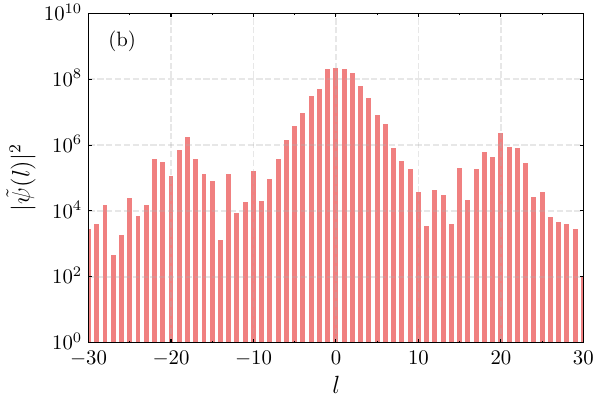} \\
\caption{Soliton: (a) Temporal evolution of density profiles of soliton, (b) OAM distribution of the soliton for $\Omega' = 0.5$, $L_{p}=1$. Here $N = 6000$, $a_s = -27.6 a_0 $, where $a_0$ is the Bohr radius, $m = 7.01$ amu,  and $P_{in} = 0.4$ pW, and all other parameters are same as in Fig.~\ref{fig:PW1}. }
  \label{fig:sol1}
\end{figure}
For this case, we consider the initial state, representing a bright soliton as

\begin{equation}
    \psi(\phi) = \sqrt{\frac{N}{\sqrt{\pi}}} e^{- \phi^2 / 2} e^{i L_p \phi}.
\end{equation}

Fig.~\ref{fig:sol1} (a) shows the non-dispersive propagation of the bright soliton within the ring structure. The slight modulation in the density profiles of soliton can be attributed to the presence of the rotating optical lattice probe. Fig.~\ref{fig:sol1} (b) shows the matter wave distribution of the solitonic state, and we find a pronounced concentration of OAM states close to $L_p=1$ and $L_p \pm 2\ell - \Omega'/2$ (with $\Omega'=0.5$). These multiple modes in the OAM distribution account for the complex internal dynamics of soliton. The occupancy in these states (corresponding to the side mode frequencies) is heightened in comparison to other states [Figs.~\ref{fig:PW1}(b) and ~\ref{fig:PW5}(b)], clearly indicating the occurrence of Bragg diffraction due to the presence of the optical lattice potential.

\begin{figure}[b]
\centering
\includegraphics[width=1\linewidth]{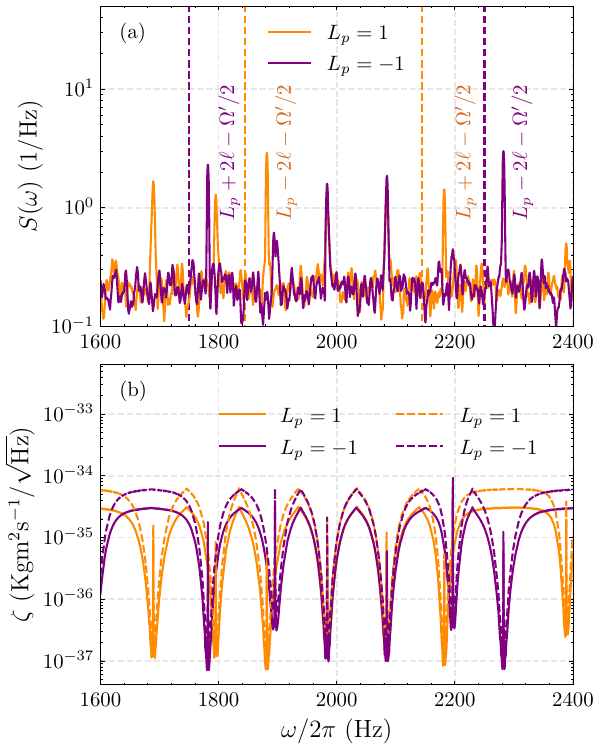} \\
\caption{Soliton: (a) Noise spectra of the output phase quadrature of the cavity field as a function of the system response frequency for $\Omega' = 0.5$, $L_{p}=1$ (orange) and $L_{p}=-1$ (purple). The vertical dashed line corresponds to the analytical predictions for the side modes of $L_p = \pm 1$ (Eq.~\ref{eq:OmP}). (b) Rotation measurement sensitivities as a function of system response frequency. The solid curves correspond to $\zeta^\Lambda$ (Eq.~\ref{eq:Sensitivity_Lp1}), while the dashed curves correspond to $\zeta^\Omega$ (Eq.~\ref{eq:Sensitivity_Lp2}). The parameters used here are the same as in Fig. \ref{fig:sol1}.}
  \label{fig:sol2}
\end{figure}

To detect the rotation of the soliton confined in the ring trap, we have calculated the noise spectra of the output phase quadrature of the cavity field for two different values of the winding number $L_p = \pm 1$ separately and have presented it as a function of the system response frequency in Fig.~\ref{fig:sol2} (a). The increased number of peaks in the output spectra, compared to the case of persistent current, accounts for the multi-mode dynamics inherent in the soliton profile. Remarkably, these two cases ($L_p = \pm 1$) yield distinct cavity output spectra, and by locating the dominating peaks in the spectrum, we can detect the sign and magnitude of the rotation of soliton in the ring. Here also, the dashed lines indicate the side mode frequencies obtained through the analytical prediction, and the numerically obtained results align closely enough with them to allow identification.

Fig.~\ref{fig:sol2} (b) shows the rotation measurement sensitivities for the two above-discussed scenarios. We can achieve the best sensitivity to the rotation measurements near the side-mode frequencies of the condensate. Along with these, we can also get better sensitivities around the frequencies corresponding to the other neighboring modes, which was not possible for the persistent current. This broader detection capability can help for a more detailed understanding of the system's behavior.


\subsubsection{Collisional dynamics of counter-propagating solitons}
\label{Soliton2}
In this section, we extend the analysis of detecting the rotation to a pair of solitons exhibiting multiple collisions~\cite{nguyen2014collisions}. These collisions depend on the effective interaction originating from the interference of two waves in the super-positioned state. We have demonstrated the situation where these collisions are repulsive by considering the initial state as
\begin{equation}
\begin{aligned}
        \psi(\phi) = \sqrt{\frac{N}{\sqrt{\pi}}} & \Big( e^{- (\phi-\pi/2)^2 / 2}  e^{i L_{p1}\phi} \\ + e^{i \theta} & e^{- (\phi+\pi/2)^2 / 2} e^{i L_{p2}\phi} \Big ).
\end{aligned}
\end{equation}


\begin{figure}[b]
\includegraphics[width=1\linewidth]{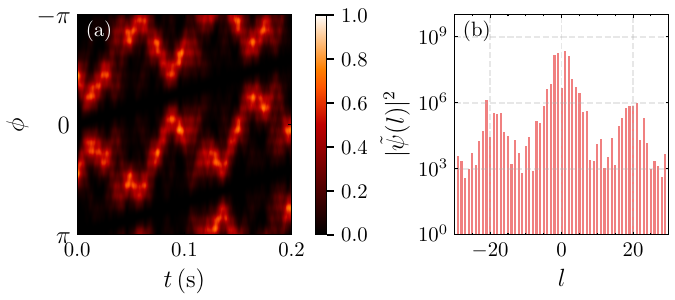} \\
\caption{Soliton collision: (a) Temporal evolution of pair of solitons showing out-of-phase collisions, (b) OAM distribution of the solitons. Here $N = 6000$, $P_{in} = 0.4$ pW, $\Omega' = 0.3$ and other parameters used are same as Fig.~\ref{fig:sol1}.}
  \label{fig:Solcol1}
\end{figure}


The above equation conveys that, initially, the pair of solitons are located at $-\pi/2$ and $\pi/2$, respectively, and the specific nature of the collision is established by setting the value of phase $\theta$ between the two solitons equal to $\pi$. Subsequently, we set up the individual soliton with distinct winding numbers $L_{p1} = -1$ and $L_{p2} = 1$, for which the solitons initially move towards each other, experiencing a repulsive collision as they move away from each other.


Fig.~\ref{fig:Solcol1} (a) shows the temporal evolution of a pair of solitons exhibiting multiple collisions over time in the ring structure in the presence of a rotating optical lattice.  Due to the phase difference of $\pi$ between the wave packets, the interference is destructive, resulting in the formation of a zero-density point. A similar phase-dependent collision of a pair of solitons was seen in the experiment of \cite{nguyen2014collisions}. Fig.~\ref{fig:Solcol1} (b) shows the OAM distribution of the condensate containing the pair of solitons, which gives information about the side mode generation as in earlier cases.

\begin{figure}[t]
\centering
\includegraphics[ width= 1\linewidth]{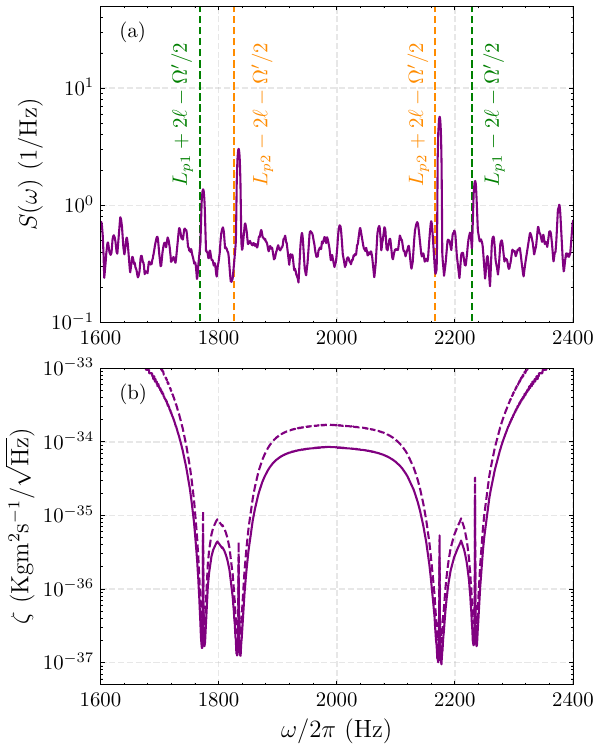} \\
\caption{Soliton collision: (a) Noise spectra of the output phase quadrature of the cavity field as a function of the system response frequency for  $\Omega' = 0.3$. The dashed lines indicate the analytical predictions for the sidemode frequencies of $L_{p1} = -1$ and $L_{p2} = 1$ respectively [Eq.~(\ref{eq:OmP})]. (b) Rotation measurement sensitivities as a function of system response frequency. The solid curves correspond to $\zeta^\Lambda$ (Eq.~\ref{eq:Sensitivity_Lp1}), while the dashed curves correspond to $\zeta^\Omega$ (Eq.~\ref{eq:Sensitivity_Lp2}). The parameters used here are the same as in Fig. \ref{fig:Solcol1}.}
\label{fig:Solcol2}
\end{figure}

\begin{figure*}[!htp]
\includegraphics[width=1\linewidth]{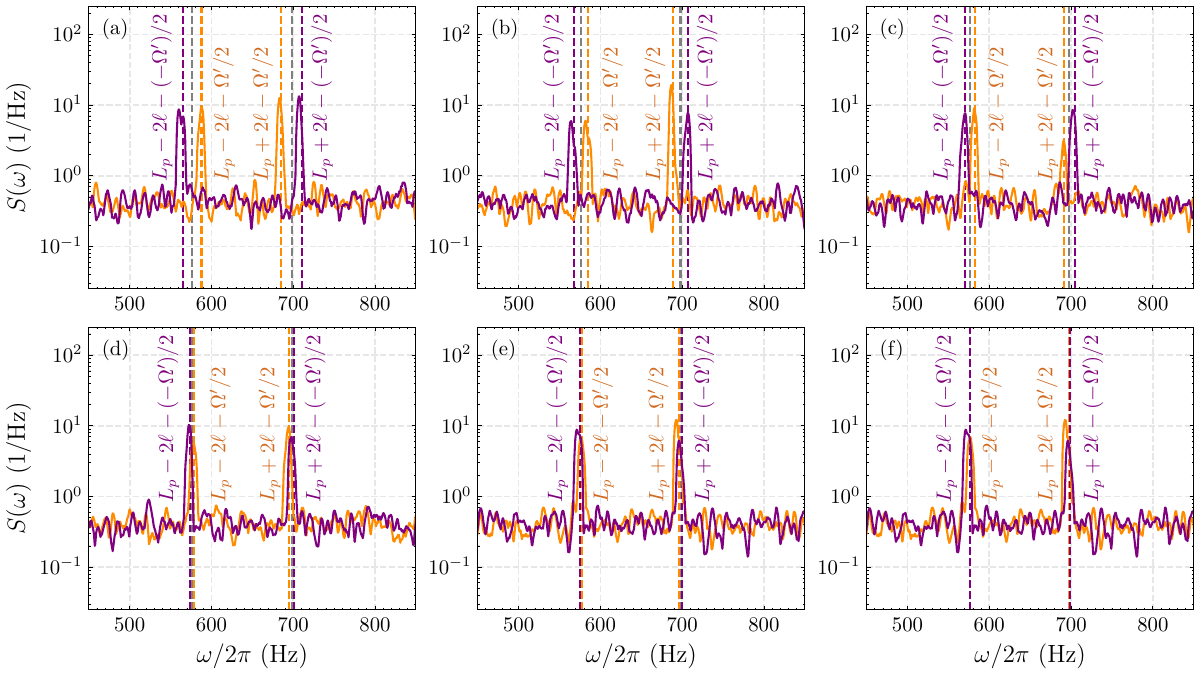} \\
\caption{Persistent current: Noise spectra of the output phase quadrature of the cavity field with $L_p = 1$ for different lattice rotation. (a) $\Omega' = 0.4$, (b) $\Omega' = 0.3$, (c) $\Omega' = 0.2$, (d) $\Omega' = 0.1$, (e) $\Omega' = 0.05$, and (f) $\Omega' = 0.01$. The gray-colored vertical dashed line corresponds to the analytical predictions for the side modes for $\Omega' = 0$ and the orange and purple-colored vertical dashed lines correspond to the analytical predictions for the side modes corresponding to the non-zero value of $\Omega'$ with $L_p = 1$ [Eq.~\ref{eq:OmP}]. The other parameters used here are the same as mentioned in Fig.~\ref{fig:PW1}. }
  \label{fig:ohm}
\end{figure*}

The noise spectrum of the phase quadrature of the cavity field is shown in Fig.~\ref{fig:Solcol2} (a). The peaks for the sidemodes corresponding to $L_{p1} = -1$ and $L_{p2} = 1$ are distinct and dominating, and yield information about the condensate winding number. The sensitivity of the rotation measurements is plotted in Fig.~\ref{fig:Solcol2} (b) as a function of the response frequency of the system and the best sensitivities are achieved near the side mode frequencies. 

\subsection{The ring BEC as a rotation sensor}
\label{sec: CritO}
In this section, we take the converse approach to that of our analysis above in order to characterize the ring BEC as a rotation sensor. We assume $L_{p}$ is known for the BEC persistent current state and investigate how the magnitude and sign of $\Omega'$ can be measured. Here $\Omega'$ now represents some unknown rotation of the laboratory that we desire to measure. As can be seen in the cavity spectra of Fig.~\ref{fig:ohm} for various $\Omega'=0.4-0.01$ for $L_p=1$, the magnitude as well as the sign of $\Omega'$ can be determined using this technique. We find that the side mode peaks corresponding to $\pm \,\Omega'$ are quite distinctly resolvable for $L_p=1$ for $\Omega'\gtrsim 0.1$. This represents the limits of the ring BEC when it is used as a laboratory rotation sensor.


\subsection{Fidelity}
\label{sec:Fidelity}
To demonstrate that our technique is minimally destructive, we have calculated the variation of fidelity of condensate wave function over time, which is defined as
\begin{equation}
\label{eq:Fidelity}
    F(t) = \int_{0}^{2\pi} \left[\psi^{*}(\phi,t)\psi(\phi,0)\right]^2 d\phi. 
\end{equation}

\begin{figure}[!htp]
\includegraphics[width=1\linewidth]{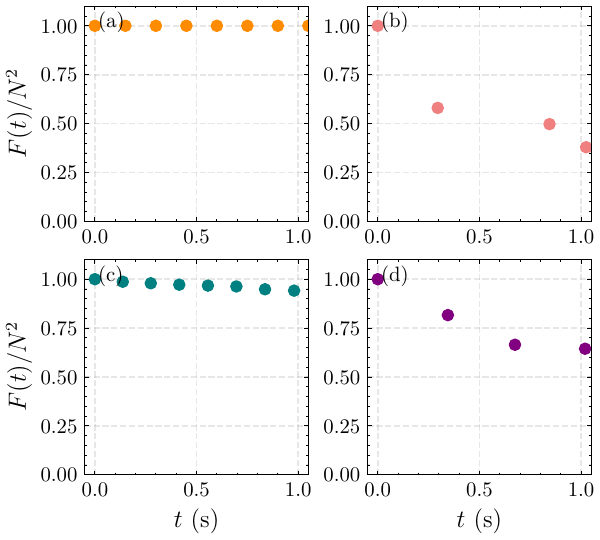} \\
\caption{Variation of fidelity with time for (a) persistent current, (b) two-state superposition of persistent current, (c) soliton, and (d) a pair of solitons. The parameters used are the same as Fig.~\ref{fig:PW1} for (a) and (b) and Fig.~\ref{fig:sol1} for (c) and (d).}
  \label{feas}
\end{figure}

Fig.~\ref{feas} depicts the variation of fidelity for the above-mentioned four cases. Fidelity remains close to unity for the persistent current case [Fig.~\ref{feas}(a)] with the slight gradual decline attributed to the measurement back action and other noises inherent to the system. In the case of soliton [Fig.~\ref{feas} (c)], the fidelity shows oscillatory behavior with time, and this is linked to the rotational motion of the soliton in the ring, particularly when the soliton's spatial position deviates from its initial state. So a meaningful calculation of fidelity occurs only at those times when the soliton realigns with the initial state, characterized by a similar density and phase distribution. At these specific times (represented by the dots), the fidelity stays close to unity, which confirms that our method is minimally destructive. 

A similar variation of fidelity emerges in the case of the superposition of persistent currents [Fig.~\ref{feas} (b)] and the pair of solitons [Fig.~\ref{feas} (d)]. However, along with the effect of the rotation, the super-positioned states are more prone to noises, as a result, the interference pattern becomes complicated. Due to this, the fidelity decreases gradually with time and the best fidelity observed, represented by the dots is near to 0.5 up to 1 sec. It should be noted that this decrease in fidelity does not imply that our method is totally destructive, which is the case for the absorption imaging technique, used in demonstrated experiments so far, for the detection of rotation
\cite{RyuPRL2007,MoulderPRA2012,WrightPRL2013,EckelNature2014, EckelPRX2018, PandeyPRL2021}.

\section{Summary and Conclusions}
\label{sec:5}
We have theoretically considered minimally destructive, \textit{in situ}, and real-time measurement of ring BEC rotation by coupling it to a cavity-carrying optical OAM. Unlike our previous proposal~\cite{KumarPRL2021,pradhan2023ringBEC}, which could only determine the magnitude and not the sign of the rotation, our present work enables the detection of both quantities. This is accomplished by 
rotating the optical lattice arising from the interference of the optical beams in the cavity.

We have analyzed the system using a few quantized light-matter modes as well as a mean-field (for the condensate) and classical (for the electromagnetic field) model. We have demonstrated the lattice rotation technique on persistent current rotational eigenstates, counter-rotating superpositions, and a soliton singly or in collision with a second soliton. Our conclusions are evidenced by numerical simulations of condensate density profiles (which characterize the measurement fidelity) and OAM content, optical transmission spectra from the cavity, and measurement sensitivities for condensate as well as lattice rotation as a function of the system response frequency. We find that the proposed technique can clearly distinguish between co- and counter-rotating excitations on the ring condensate. The predictions of the few-mode and multi-mode models are in good agreement with each other. We have also found the critical lattice rotation frequency above which our proposed technique is clearly able to lattice rotation, indicating the use of the ring BEC as a rotation sensor.

We expect the technique proposed by us to be of interest in the context of a wide class of experiments where the measurement of ring condensate rotation is of interest, such as superfluid hydrodynamics, atomtronics, and soliton interferometry, as well as for rotation sensing. 
\section{Acknowledgments}
We thank the International Centre for Theoretical Sciences, Bengaluru, where this work was initiated, for hosting us. M.B. would like to thank the Air Force Office of Scientific Research (FA9550-23-1-0259) for support. R.K. acknowledges support from JSPS KAKENHI Grant No. JP21K03421. P.K. acknowledges the financial support from the Max Planck Society. We also gratefully acknowledge our super-computing facility Param-Ishan (IITG), where all the simulation runs were performed.
\newpage
\appendix

\counterwithin{figure}{section}

\section{Chirality detection and rotation sensing at high rotation rate of optical lattice}
\label{App:highRot}
In this appendix, we present the simulation results pertaining to the chirality detection and rotational sensitivity measurement for the situation when the condensate is subject to high rotation along with the optical trap. 

\subsection{Persistent current} 
\begin{figure*}[!htb]
\centering
\includegraphics[width=1.05\linewidth]{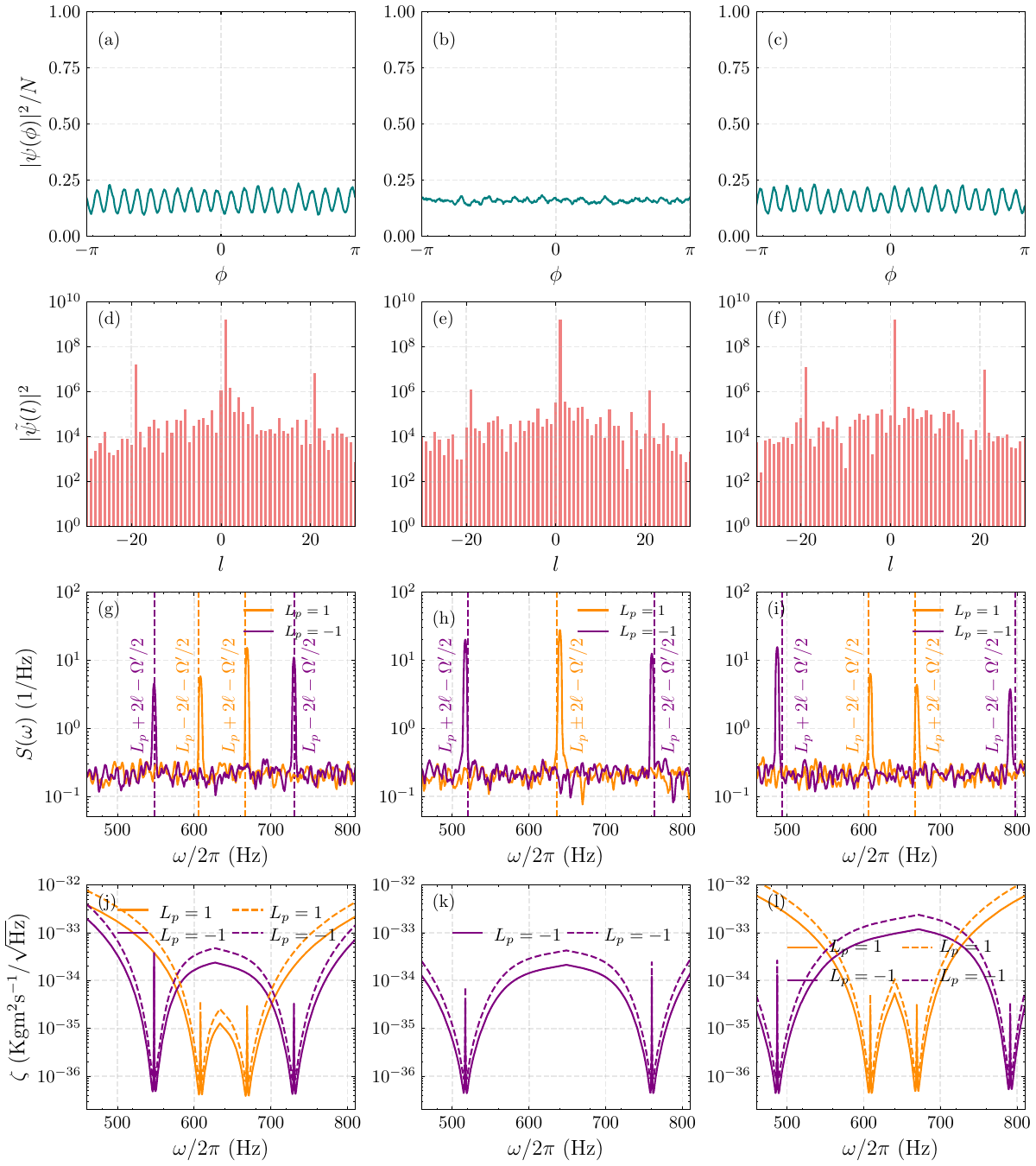}

\caption{Persistent current: (a)-(c) Density profiles of persistent current for $L_p = 1$, (d)-(f) OAM states of the condensate for $L_p = 1$, (g)-(i) Noise spectrum of the output phase quadrature versus response frequency. The vertical dashed line corresponds to the analytical predictions for the side modes (Colors are used in the same way as in Fig: \ref{fig:PW2}). (j)-(l) Rotation measurement sensitivity as a function of system response frequency. The solid curves correspond to $\zeta^\Lambda$ (Eq.~\ref{eq:Sensitivity_Lp1}), while the dashed curves correspond to $\zeta^\Omega$ (Eq.~\ref{eq:Sensitivity_Lp2}).  Here $P_{in} = 50$ fW and $\Omega' = 1$,  $2$, and $3$ respectively. The remaining set of parameters are the same as in Fig.~\ref{fig:PW1}. }
    \label{fig:PW4}
\end{figure*}

The corresponding results for persistent current are shown in Figs.~\ref{fig:PW4}.  (a) - (c) depicts the condensate density profiles for increasing the rotation frequency of the optical lattice for $L_p = 1$. We note that for $\Omega' = 2 L_p$, the side mode frequencies $\omega'_{c} \equiv \omega_{d}' = \pm 2\ell$ [Fig.~\ref{fig:PW4} (b)]. The corresponding OAM distributions are presented in Figs.~\ref{fig:PW4} (d) - (f). When $\Omega' = 2$ for $L_p = 1$ [Fig.~\ref{fig:PW4} (e)], the occupancies of the $L_p \pm 2\ell$ modes are nearly equal and decreased by an order of magnitude from the other cases. 

Figs.~\ref{fig:PW4} (g) - (i) show the noise spectra of the phase quadrature of the cavity transmission for the above-mentioned three cases $\Omega' = 1,2$ and $3$, respectively. When $\Omega'/2 \neq L_p$, we get distinct peaks for $L_p = 1$ and $L_p = -1$ and the information about the winding number can be detected. Otherwise, we obtain a single degenerate peak in the output spectrum (when $\Omega'/2 = L_p $) and by using this relation between $\Omega'$ and $L_p$, the magnitude and sign of the winding number of the persistent current can be obtained. The corresponding rotation measurement sensitivities are shown in Figs.~\ref{fig:PW4} (j) - (l) and the best sensitivities are obtained near the side mode frequencies as in earlier cases. 

\subsection{Soliton} 
\begin{figure*}[!htb]
\centering
\includegraphics[width=1\linewidth]{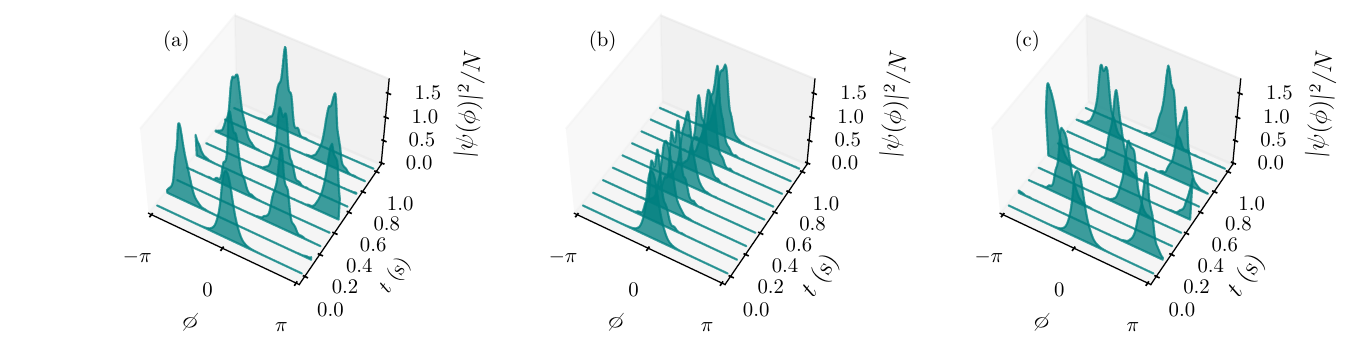}
\includegraphics[width=1\linewidth]{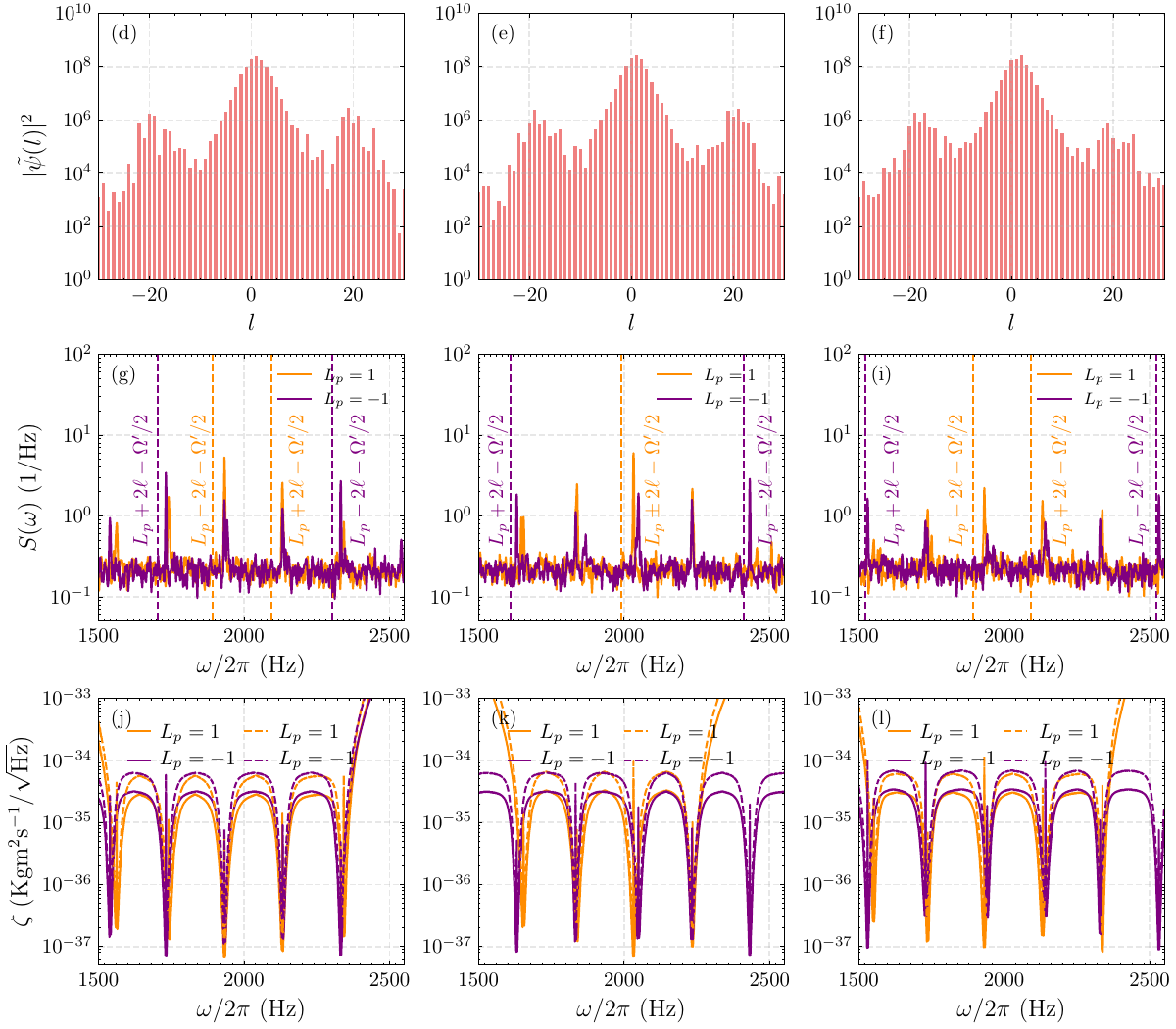}

\caption{Soliton: (a)-(c) Density profiles of soliton for $L_p = 1$, (d)-(f) OAM states of the condensate for $L_p = 1$, (g)-(i) Power spectrum of the imaginary part of cavity field versus response frequency. The vertical dashed line corresponds to the analytical predictions for the side modes (Colors are used in the same way as in Fig.~\ref{fig:PW2}). (j)-(l) Rotation measurement sensitivities as a function of system response frequency. The solid curves correspond to $\zeta^\Lambda$ (Eq.~\ref{eq:Sensitivity_Lp1}), while the dashed curves correspond to $\zeta^\Omega$ (Eq.~\ref{eq:Sensitivity_Lp2}).  Here $P_{in} = 0.4$ pW and $\Omega' = 1$,  $2$, and $3$ respectively. The remaining set of parameters are the same as in Fig.~\ref{fig:sol1}.}
    \label{fig:sol3}
\end{figure*}

The rotation of soliton in the ring structure is presented in Figs.~\ref{fig:sol3} (a) - (c), corresponding to $\Omega' = 1$,  $2$, and $3$, respectively, and the corresponding orbital angular momentum distribution are shown in (d) - (f). The noise spectra of the phase quadrature of the cavity field for these three cases are shown in Figs.~\ref{fig:sol3} (g) - (i), respectively. Due to the multi-mode nature of the soliton, a larger number of peaks is observed, as discussed earlier in Section \ref{sec: Soliton}; nonetheless, by locating and measuring the dominating peaks, the information about the winding number can be extracted accurately. 


\clearpage
\bibliography{citation.bib} 

\begin{thebibliography}{61}%
\makeatletter
\providecommand \@ifxundefined [1]{%
 \@ifx{#1\undefined}
}%
\providecommand \@ifnum [1]{%
 \ifnum #1\expandafter \@firstoftwo
 \else \expandafter \@secondoftwo
 \fi
}%
\providecommand \@ifx [1]{%
 \ifx #1\expandafter \@firstoftwo
 \else \expandafter \@secondoftwo
 \fi
}%
\providecommand \natexlab [1]{#1}%
\providecommand \enquote  [1]{``#1''}%
\providecommand \bibnamefont  [1]{#1}%
\providecommand \bibfnamefont [1]{#1}%
\providecommand \citenamefont [1]{#1}%
\providecommand \href@noop [0]{\@secondoftwo}%
\providecommand \href [0]{\begingroup \@sanitize@url \@href}%
\providecommand \@href[1]{\@@startlink{#1}\@@href}%
\providecommand \@@href[1]{\endgroup#1\@@endlink}%
\providecommand \@sanitize@url [0]{\catcode `\\12\catcode `\$12\catcode
  `\&12\catcode `\#12\catcode `\^12\catcode `\_12\catcode `\%12\relax}%
\providecommand \@@startlink[1]{}%
\providecommand \@@endlink[0]{}%
\providecommand \url  [0]{\begingroup\@sanitize@url \@url }%
\providecommand \@url [1]{\endgroup\@href {#1}{\urlprefix }}%
\providecommand \urlprefix  [0]{URL }%
\providecommand \Eprint [0]{\href }%
\providecommand \doibase [0]{https://doi.org/}%
\providecommand \selectlanguage [0]{\@gobble}%
\providecommand \bibinfo  [0]{\@secondoftwo}%
\providecommand \bibfield  [0]{\@secondoftwo}%
\providecommand \translation [1]{[#1]}%
\providecommand \BibitemOpen [0]{}%
\providecommand \bibitemStop [0]{}%
\providecommand \bibitemNoStop [0]{.\EOS\space}%
\providecommand \EOS [0]{\spacefactor3000\relax}%
\providecommand \BibitemShut  [1]{\csname bibitem#1\endcsname}%
\let\auto@bib@innerbib\@empty
\bibitem [{\citenamefont {Kumar}\ \emph {et~al.}(2021)\citenamefont {Kumar},
  \citenamefont {Biswas}, \citenamefont {Feliz}, \citenamefont {Kanamoto},
  \citenamefont {Chang}, \citenamefont {Jha},\ and\ \citenamefont
  {Bhattacharya}}]{KumarPRL2021}%
  \BibitemOpen
  \bibfield  {author} {\bibinfo {author} {\bibfnamefont {P.}~\bibnamefont
  {Kumar}}, \bibinfo {author} {\bibfnamefont {T.}~\bibnamefont {Biswas}},
  \bibinfo {author} {\bibfnamefont {K.}~\bibnamefont {Feliz}}, \bibinfo
  {author} {\bibfnamefont {R.}~\bibnamefont {Kanamoto}}, \bibinfo {author}
  {\bibfnamefont {M.-S.}\ \bibnamefont {Chang}}, \bibinfo {author}
  {\bibfnamefont {A.~K.}\ \bibnamefont {Jha}},\ and\ \bibinfo {author}
  {\bibfnamefont {M.}~\bibnamefont {Bhattacharya}},\ }\bibfield  {title}
  {\bibinfo {title} {{Cavity Optomechanical Sensing and Manipulation of an
  Atomic Persistent Current}},\ }\href
  {https://doi.org/10.1103/PhysRevLett.127.113601} {\bibfield  {journal}
  {\bibinfo  {journal} {Phys. Rev. Lett.}\ }\textbf {\bibinfo {volume} {127}},\
  \bibinfo {pages} {113601} (\bibinfo {year} {2021})}\BibitemShut {NoStop}%
\bibitem [{\citenamefont {Pradhan}\ \emph {et~al.}(2023)\citenamefont
  {Pradhan}, \citenamefont {Kumar}, \citenamefont {Kanamoto}, \citenamefont
  {Dey}, \citenamefont {Bhattacharya},\ and\ \citenamefont
  {Mishra}}]{pradhan2023ringBEC}%
  \BibitemOpen
  \bibfield  {author} {\bibinfo {author} {\bibfnamefont {N.}~\bibnamefont
  {Pradhan}}, \bibinfo {author} {\bibfnamefont {P.}~\bibnamefont {Kumar}},
  \bibinfo {author} {\bibfnamefont {R.}~\bibnamefont {Kanamoto}}, \bibinfo
  {author} {\bibfnamefont {T.~N.}\ \bibnamefont {Dey}}, \bibinfo {author}
  {\bibfnamefont {M.}~\bibnamefont {Bhattacharya}},\ and\ \bibinfo {author}
  {\bibfnamefont {P.~K.}\ \bibnamefont {Mishra}},\ }\bibfield  {title}
  {\bibinfo {title} {{Cavity optomechanical detection of persistent currents
  and solitons in a bosonic ring condensate}},\ }\href@noop {} {\bibfield
  {journal} {\bibinfo  {journal} {arXiv preprint arXiv:2306.06720}\ } (\bibinfo
  {year} {2023})}\BibitemShut {NoStop}%
\bibitem [{\citenamefont {Ryu}\ \emph {et~al.}(2007)\citenamefont {Ryu},
  \citenamefont {Andersen}, \citenamefont {Clad\'e}, \citenamefont {Natarajan},
  \citenamefont {Helmerson},\ and\ \citenamefont {Phillips}}]{RyuPRL2007}%
  \BibitemOpen
  \bibfield  {author} {\bibinfo {author} {\bibfnamefont {C.}~\bibnamefont
  {Ryu}}, \bibinfo {author} {\bibfnamefont {M.~F.}\ \bibnamefont {Andersen}},
  \bibinfo {author} {\bibfnamefont {P.}~\bibnamefont {Clad\'e}}, \bibinfo
  {author} {\bibfnamefont {V.}~\bibnamefont {Natarajan}}, \bibinfo {author}
  {\bibfnamefont {K.}~\bibnamefont {Helmerson}},\ and\ \bibinfo {author}
  {\bibfnamefont {W.~D.}\ \bibnamefont {Phillips}},\ }\bibfield  {title}
  {\bibinfo {title} {{Observation of Persistent Flow of a Bose-Einstein
  Condensate in a Toroidal Trap}},\ }\href
  {https://doi.org/10.1103/PhysRevLett.99.260401} {\bibfield  {journal}
  {\bibinfo  {journal} {Phys. Rev. Lett.}\ }\textbf {\bibinfo {volume} {99}},\
  \bibinfo {pages} {260401} (\bibinfo {year} {2007})}\BibitemShut {NoStop}%
\bibitem [{\citenamefont {Guo}\ \emph {et~al.}(2020)\citenamefont {Guo},
  \citenamefont {Dubessy}, \citenamefont {de~Herve}, \citenamefont {Kumar},
  \citenamefont {Badr}, \citenamefont {Perrin}, \citenamefont {Longchambon},\
  and\ \citenamefont {Perrin}}]{GuoPRL2020}%
  \BibitemOpen
  \bibfield  {author} {\bibinfo {author} {\bibfnamefont {Y.}~\bibnamefont
  {Guo}}, \bibinfo {author} {\bibfnamefont {R.}~\bibnamefont {Dubessy}},
  \bibinfo {author} {\bibfnamefont {M.~d.~G.}\ \bibnamefont {de~Herve}},
  \bibinfo {author} {\bibfnamefont {A.}~\bibnamefont {Kumar}}, \bibinfo
  {author} {\bibfnamefont {T.}~\bibnamefont {Badr}}, \bibinfo {author}
  {\bibfnamefont {A.}~\bibnamefont {Perrin}}, \bibinfo {author} {\bibfnamefont
  {L.}~\bibnamefont {Longchambon}},\ and\ \bibinfo {author} {\bibfnamefont
  {H.}~\bibnamefont {Perrin}},\ }\bibfield  {title} {\bibinfo {title}
  {Supersonic rotation of a superfluid: A long-lived dynamical ring},\ }\href
  {https://doi.org/10.1103/PhysRevLett.124.025301} {\bibfield  {journal}
  {\bibinfo  {journal} {Phys. Rev. Lett.}\ }\textbf {\bibinfo {volume} {124}},\
  \bibinfo {pages} {025301} (\bibinfo {year} {2020})}\BibitemShut {NoStop}%
\bibitem [{\citenamefont {Cai}\ \emph {et~al.}(2022)\citenamefont {Cai},
  \citenamefont {Allman}, \citenamefont {Sabharwal},\ and\ \citenamefont
  {Wright}}]{CaiPRL2022}%
  \BibitemOpen
  \bibfield  {author} {\bibinfo {author} {\bibfnamefont {Y.}~\bibnamefont
  {Cai}}, \bibinfo {author} {\bibfnamefont {D.~G.}\ \bibnamefont {Allman}},
  \bibinfo {author} {\bibfnamefont {P.}~\bibnamefont {Sabharwal}},\ and\
  \bibinfo {author} {\bibfnamefont {K.~C.}\ \bibnamefont {Wright}},\ }\bibfield
   {title} {\bibinfo {title} {Persistent currents in rings of ultracold
  fermionic atoms},\ }\href {https://doi.org/10.1103/PhysRevLett.128.150401}
  {\bibfield  {journal} {\bibinfo  {journal} {Phys. Rev. Lett.}\ }\textbf
  {\bibinfo {volume} {128}},\ \bibinfo {pages} {150401} (\bibinfo {year}
  {2022})}\BibitemShut {NoStop}%
\bibitem [{\citenamefont {Del~Pace}\ \emph {et~al.}(2022)\citenamefont
  {Del~Pace}, \citenamefont {Xhani}, \citenamefont {Muzi~Falconi},
  \citenamefont {Fedrizzi}, \citenamefont {Grani}, \citenamefont
  {Hernandez~Rajkov}, \citenamefont {Inguscio}, \citenamefont {Scazza},
  \citenamefont {Kwon},\ and\ \citenamefont {Roati}}]{DelPacePRX2022}%
  \BibitemOpen
  \bibfield  {author} {\bibinfo {author} {\bibfnamefont {G.}~\bibnamefont
  {Del~Pace}}, \bibinfo {author} {\bibfnamefont {K.}~\bibnamefont {Xhani}},
  \bibinfo {author} {\bibfnamefont {A.}~\bibnamefont {Muzi~Falconi}}, \bibinfo
  {author} {\bibfnamefont {M.}~\bibnamefont {Fedrizzi}}, \bibinfo {author}
  {\bibfnamefont {N.}~\bibnamefont {Grani}}, \bibinfo {author} {\bibfnamefont
  {D.}~\bibnamefont {Hernandez~Rajkov}}, \bibinfo {author} {\bibfnamefont
  {M.}~\bibnamefont {Inguscio}}, \bibinfo {author} {\bibfnamefont
  {F.}~\bibnamefont {Scazza}}, \bibinfo {author} {\bibfnamefont {W.~J.}\
  \bibnamefont {Kwon}},\ and\ \bibinfo {author} {\bibfnamefont
  {G.}~\bibnamefont {Roati}},\ }\bibfield  {title} {\bibinfo {title}
  {Imprinting persistent currents in tunable fermionic rings},\ }\href
  {https://doi.org/10.1103/PhysRevX.12.041037} {\bibfield  {journal} {\bibinfo
  {journal} {Phys. Rev. X}\ }\textbf {\bibinfo {volume} {12}},\ \bibinfo
  {pages} {041037} (\bibinfo {year} {2022})}\BibitemShut {NoStop}%
\bibitem [{\citenamefont {Guo}\ \emph {et~al.}(2022)\citenamefont {Guo},
  \citenamefont {Gutierrez}, \citenamefont {Rey}, \citenamefont {Badr},
  \citenamefont {Perrin}, \citenamefont {Longchambon}, \citenamefont {Bagnato},
  \citenamefont {Perrin},\ and\ \citenamefont {Dubessy}}]{GuoNJP2022}%
  \BibitemOpen
  \bibfield  {author} {\bibinfo {author} {\bibfnamefont {Y.}~\bibnamefont
  {Guo}}, \bibinfo {author} {\bibfnamefont {E.~M.}\ \bibnamefont {Gutierrez}},
  \bibinfo {author} {\bibfnamefont {D.}~\bibnamefont {Rey}}, \bibinfo {author}
  {\bibfnamefont {T.}~\bibnamefont {Badr}}, \bibinfo {author} {\bibfnamefont
  {T.}~\bibnamefont {Perrin}}, \bibinfo {author} {\bibfnamefont
  {L.}~\bibnamefont {Longchambon}}, \bibinfo {author} {\bibfnamefont {V.~S.}\
  \bibnamefont {Bagnato}}, \bibinfo {author} {\bibfnamefont {H.}~\bibnamefont
  {Perrin}},\ and\ \bibinfo {author} {\bibfnamefont {R.}~\bibnamefont
  {Dubessy}},\ }\bibfield  {title} {\bibinfo {title} {Expansion of a quantum
  gas in a shell trap},\ }\href {https://doi.org/10.1088/1367-2630/ac919f}
  {\bibfield  {journal} {\bibinfo  {journal} {New J. Phys.}\ }\textbf {\bibinfo
  {volume} {24}},\ \bibinfo {pages} {093040} (\bibinfo {year}
  {2022})}\BibitemShut {NoStop}%
\bibitem [{\citenamefont {Moulder}\ \emph {et~al.}(2012)\citenamefont
  {Moulder}, \citenamefont {Beattie}, \citenamefont {Smith}, \citenamefont
  {Tammuz},\ and\ \citenamefont {Hadzibabic}}]{MoulderPRA2012}%
  \BibitemOpen
  \bibfield  {author} {\bibinfo {author} {\bibfnamefont {S.}~\bibnamefont
  {Moulder}}, \bibinfo {author} {\bibfnamefont {S.}~\bibnamefont {Beattie}},
  \bibinfo {author} {\bibfnamefont {R.~P.}\ \bibnamefont {Smith}}, \bibinfo
  {author} {\bibfnamefont {N.}~\bibnamefont {Tammuz}},\ and\ \bibinfo {author}
  {\bibfnamefont {Z.}~\bibnamefont {Hadzibabic}},\ }\bibfield  {title}
  {\bibinfo {title} {{Quantized supercurrent decay in an annular Bose-Einstein
  condensate}},\ }\href {https://doi.org/10.1103/PhysRevA.86.013629} {\bibfield
   {journal} {\bibinfo  {journal} {Phys. Rev. A}\ }\textbf {\bibinfo {volume}
  {86}},\ \bibinfo {pages} {013629} (\bibinfo {year} {2012})}\BibitemShut
  {NoStop}%
\bibitem [{\citenamefont {Wright}\ \emph
  {et~al.}(2013{\natexlab{a}})\citenamefont {Wright}, \citenamefont
  {Blakestad}, \citenamefont {Lobb}, \citenamefont {Phillips},\ and\
  \citenamefont {Campbell}}]{WrightPRL2013}%
  \BibitemOpen
  \bibfield  {author} {\bibinfo {author} {\bibfnamefont {K.~C.}\ \bibnamefont
  {Wright}}, \bibinfo {author} {\bibfnamefont {R.~B.}\ \bibnamefont
  {Blakestad}}, \bibinfo {author} {\bibfnamefont {C.~J.}\ \bibnamefont {Lobb}},
  \bibinfo {author} {\bibfnamefont {W.~D.}\ \bibnamefont {Phillips}},\ and\
  \bibinfo {author} {\bibfnamefont {G.~K.}\ \bibnamefont {Campbell}},\
  }\bibfield  {title} {\bibinfo {title} {{Driving Phase Slips in a Superfluid
  Atom Circuit with a Rotating Weak Link}},\ }\href
  {https://doi.org/10.1103/PhysRevLett.110.025302} {\bibfield  {journal}
  {\bibinfo  {journal} {Phys. Rev. Lett.}\ }\textbf {\bibinfo {volume} {110}},\
  \bibinfo {pages} {025302} (\bibinfo {year} {2013}{\natexlab{a}})}\BibitemShut
  {NoStop}%
\bibitem [{\citenamefont {Snizhko}\ \emph {et~al.}(2016)\citenamefont
  {Snizhko}, \citenamefont {Isaieva}, \citenamefont {Kuriatnikov},
  \citenamefont {Bidasyuk}, \citenamefont {Vilchinskii},\ and\ \citenamefont
  {Yakimenko}}]{SnizhkoPRA2016}%
  \BibitemOpen
  \bibfield  {author} {\bibinfo {author} {\bibfnamefont {K.}~\bibnamefont
  {Snizhko}}, \bibinfo {author} {\bibfnamefont {K.}~\bibnamefont {Isaieva}},
  \bibinfo {author} {\bibfnamefont {Y.}~\bibnamefont {Kuriatnikov}}, \bibinfo
  {author} {\bibfnamefont {Y.}~\bibnamefont {Bidasyuk}}, \bibinfo {author}
  {\bibfnamefont {S.}~\bibnamefont {Vilchinskii}},\ and\ \bibinfo {author}
  {\bibfnamefont {A.}~\bibnamefont {Yakimenko}},\ }\bibfield  {title} {\bibinfo
  {title} {{Stochastic phase slips in toroidal Bose-Einstein condensates}},\
  }\href {https://doi.org/10.1103/PhysRevA.94.063642} {\bibfield  {journal}
  {\bibinfo  {journal} {Phys. Rev. A}\ }\textbf {\bibinfo {volume} {94}},\
  \bibinfo {pages} {063642} (\bibinfo {year} {2016})}\BibitemShut {NoStop}%
\bibitem [{\citenamefont {Kanamoto}\ \emph {et~al.}(2003)\citenamefont
  {Kanamoto}, \citenamefont {Saito},\ and\ \citenamefont
  {Ueda}}]{KanamotoPRA2003}%
  \BibitemOpen
  \bibfield  {author} {\bibinfo {author} {\bibfnamefont {R.}~\bibnamefont
  {Kanamoto}}, \bibinfo {author} {\bibfnamefont {H.}~\bibnamefont {Saito}},\
  and\ \bibinfo {author} {\bibfnamefont {M.}~\bibnamefont {Ueda}},\ }\bibfield
  {title} {\bibinfo {title} {{Stability of the quantized circulation of an
  attractive Bose-Einstein condensate in a rotating torus}},\ }\href
  {https://doi.org/10.1103/PhysRevA.68.043619} {\bibfield  {journal} {\bibinfo
  {journal} {Phys. Rev. A}\ }\textbf {\bibinfo {volume} {68}},\ \bibinfo
  {pages} {043619} (\bibinfo {year} {2003})}\BibitemShut {NoStop}%
\bibitem [{\citenamefont {Eckel}\ \emph {et~al.}(2014)\citenamefont {Eckel},
  \citenamefont {Lee}, \citenamefont {Jendrzejewski}, \citenamefont {Murray},
  \citenamefont {Clark}, \citenamefont {Lobb}, \citenamefont {Phillips},
  \citenamefont {Edwards},\ and\ \citenamefont {Campbell}}]{EckelNature2014}%
  \BibitemOpen
  \bibfield  {author} {\bibinfo {author} {\bibfnamefont {S.}~\bibnamefont
  {Eckel}}, \bibinfo {author} {\bibfnamefont {J.}~\bibnamefont {Lee}}, \bibinfo
  {author} {\bibfnamefont {F.}~\bibnamefont {Jendrzejewski}}, \bibinfo {author}
  {\bibfnamefont {N.}~\bibnamefont {Murray}}, \bibinfo {author} {\bibfnamefont
  {C.~W.}\ \bibnamefont {Clark}}, \bibinfo {author} {\bibfnamefont {C.~J.}\
  \bibnamefont {Lobb}}, \bibinfo {author} {\bibfnamefont {W.~D.}\ \bibnamefont
  {Phillips}}, \bibinfo {author} {\bibfnamefont {M.}~\bibnamefont {Edwards}},\
  and\ \bibinfo {author} {\bibfnamefont {G.~K.}\ \bibnamefont {Campbell}},\
  }\bibfield  {title} {\bibinfo {title} {Hysteresis in a quantized superfluid
  ‘atomtronic’ circuit},\ }\href
  {https://doi.org/https://doi.org/10.1038/nature12958} {\bibfield  {journal}
  {\bibinfo  {journal} {Nature}\ }\textbf {\bibinfo {volume} {506}},\ \bibinfo
  {pages} {200} (\bibinfo {year} {2014})}\BibitemShut {NoStop}%
\bibitem [{\citenamefont {Hou}\ \emph {et~al.}(2017)\citenamefont {Hou},
  \citenamefont {Luo}, \citenamefont {Sun},\ and\ \citenamefont
  {Zhang}}]{HouPRA2017}%
  \BibitemOpen
  \bibfield  {author} {\bibinfo {author} {\bibfnamefont {J.}~\bibnamefont
  {Hou}}, \bibinfo {author} {\bibfnamefont {X.-W.}\ \bibnamefont {Luo}},
  \bibinfo {author} {\bibfnamefont {K.}~\bibnamefont {Sun}},\ and\ \bibinfo
  {author} {\bibfnamefont {C.}~\bibnamefont {Zhang}},\ }\bibfield  {title}
  {\bibinfo {title} {{Adiabatically tuning quantized supercurrents in an
  annular Bose-Einstein condensate}},\ }\href
  {https://doi.org/10.1103/PhysRevA.96.011603} {\bibfield  {journal} {\bibinfo
  {journal} {Phys. Rev. A}\ }\textbf {\bibinfo {volume} {96}},\ \bibinfo
  {pages} {011603} (\bibinfo {year} {2017})}\BibitemShut {NoStop}%
\bibitem [{\citenamefont {Wright}\ \emph
  {et~al.}(2013{\natexlab{b}})\citenamefont {Wright}, \citenamefont
  {Blakestad}, \citenamefont {Lobb}, \citenamefont {Phillips},\ and\
  \citenamefont {Campbell}}]{WrightPRA2013}%
  \BibitemOpen
  \bibfield  {author} {\bibinfo {author} {\bibfnamefont {K.~C.}\ \bibnamefont
  {Wright}}, \bibinfo {author} {\bibfnamefont {R.~B.}\ \bibnamefont
  {Blakestad}}, \bibinfo {author} {\bibfnamefont {C.~J.}\ \bibnamefont {Lobb}},
  \bibinfo {author} {\bibfnamefont {W.~D.}\ \bibnamefont {Phillips}},\ and\
  \bibinfo {author} {\bibfnamefont {G.~K.}\ \bibnamefont {Campbell}},\
  }\bibfield  {title} {\bibinfo {title} {Threshold for creating excitations in
  a stirred superfluid ring},\ }\href
  {https://doi.org/10.1103/PhysRevA.88.063633} {\bibfield  {journal} {\bibinfo
  {journal} {Phys. Rev. A}\ }\textbf {\bibinfo {volume} {88}},\ \bibinfo
  {pages} {063633} (\bibinfo {year} {2013}{\natexlab{b}})}\BibitemShut
  {NoStop}%
\bibitem [{\citenamefont {Beattie}\ \emph {et~al.}(2013)\citenamefont
  {Beattie}, \citenamefont {Moulder}, \citenamefont {Fletcher},\ and\
  \citenamefont {Hadzibabic}}]{BeattiePRL2013}%
  \BibitemOpen
  \bibfield  {author} {\bibinfo {author} {\bibfnamefont {S.}~\bibnamefont
  {Beattie}}, \bibinfo {author} {\bibfnamefont {S.}~\bibnamefont {Moulder}},
  \bibinfo {author} {\bibfnamefont {R.~J.}\ \bibnamefont {Fletcher}},\ and\
  \bibinfo {author} {\bibfnamefont {Z.}~\bibnamefont {Hadzibabic}},\ }\bibfield
   {title} {\bibinfo {title} {{Persistent Currents in Spinor Condensates}},\
  }\href {https://doi.org/10.1103/PhysRevLett.110.025301} {\bibfield  {journal}
  {\bibinfo  {journal} {Phys. Rev. Lett.}\ }\textbf {\bibinfo {volume} {110}},\
  \bibinfo {pages} {025301} (\bibinfo {year} {2013})}\BibitemShut {NoStop}%
\bibitem [{\citenamefont {Gallemí}\ \emph {et~al.}(2015)\citenamefont
  {Gallemí}, \citenamefont {Mateo}, \citenamefont {Mayol},\ and\ \citenamefont
  {Guilleumas}}]{GallemiNJP2015}%
  \BibitemOpen
  \bibfield  {author} {\bibinfo {author} {\bibfnamefont {A.}~\bibnamefont
  {Gallemí}}, \bibinfo {author} {\bibfnamefont {A.~M.}\ \bibnamefont {Mateo}},
  \bibinfo {author} {\bibfnamefont {R.}~\bibnamefont {Mayol}},\ and\ \bibinfo
  {author} {\bibfnamefont {M.}~\bibnamefont {Guilleumas}},\ }\bibfield  {title}
  {\bibinfo {title} {Coherent quantum phase slip in two-component bosonic
  atomtronic circuits},\ }\href {https://doi.org/10.1088/1367-2630/18/1/015003}
  {\bibfield  {journal} {\bibinfo  {journal} {New J. Phys.}\ }\textbf {\bibinfo
  {volume} {18}},\ \bibinfo {pages} {015003} (\bibinfo {year}
  {2015})}\BibitemShut {NoStop}%
\bibitem [{\citenamefont {Wang}\ \emph {et~al.}(2015)\citenamefont {Wang},
  \citenamefont {Kumar}, \citenamefont {Jendrzejewski}, \citenamefont {Wilson},
  \citenamefont {Edwards}, \citenamefont {Eckel}, \citenamefont {Campbell},\
  and\ \citenamefont {Clark}}]{WangNJP2015}%
  \BibitemOpen
  \bibfield  {author} {\bibinfo {author} {\bibfnamefont {Y.~H.}\ \bibnamefont
  {Wang}}, \bibinfo {author} {\bibfnamefont {A.}~\bibnamefont {Kumar}},
  \bibinfo {author} {\bibfnamefont {F.}~\bibnamefont {Jendrzejewski}}, \bibinfo
  {author} {\bibfnamefont {R.~M.}\ \bibnamefont {Wilson}}, \bibinfo {author}
  {\bibfnamefont {M.}~\bibnamefont {Edwards}}, \bibinfo {author} {\bibfnamefont
  {S.}~\bibnamefont {Eckel}}, \bibinfo {author} {\bibfnamefont {G.~K.}\
  \bibnamefont {Campbell}},\ and\ \bibinfo {author} {\bibfnamefont {C.~W.}\
  \bibnamefont {Clark}},\ }\bibfield  {title} {\bibinfo {title} {{Resonant
  wavepackets and shock waves in an atomtronic SQUID}},\ }\href
  {https://doi.org/10.1088/1367-2630/17/12/125012} {\bibfield  {journal}
  {\bibinfo  {journal} {New J. Phys.}\ }\textbf {\bibinfo {volume} {17}},\
  \bibinfo {pages} {125012} (\bibinfo {year} {2015})}\BibitemShut {NoStop}%
\bibitem [{\citenamefont {Kanamoto}\ \emph {et~al.}(2008)\citenamefont
  {Kanamoto}, \citenamefont {Carr},\ and\ \citenamefont
  {Ueda}}]{KanamotoPRL2008}%
  \BibitemOpen
  \bibfield  {author} {\bibinfo {author} {\bibfnamefont {R.}~\bibnamefont
  {Kanamoto}}, \bibinfo {author} {\bibfnamefont {L.~D.}\ \bibnamefont {Carr}},\
  and\ \bibinfo {author} {\bibfnamefont {M.}~\bibnamefont {Ueda}},\ }\bibfield
  {title} {\bibinfo {title} {{Topological Winding and Unwinding in Metastable
  Bose-Einstein Condensates}},\ }\href
  {https://doi.org/10.1103/PhysRevLett.100.060401} {\bibfield  {journal}
  {\bibinfo  {journal} {Phys. Rev. Lett.}\ }\textbf {\bibinfo {volume} {100}},\
  \bibinfo {pages} {060401} (\bibinfo {year} {2008})}\BibitemShut {NoStop}%
\bibitem [{\citenamefont {Das}\ \emph {et~al.}(2012{\natexlab{a}})\citenamefont
  {Das}, \citenamefont {Sabbatini},\ and\ \citenamefont
  {Zurek}}]{DasScientificReports2012}%
  \BibitemOpen
  \bibfield  {author} {\bibinfo {author} {\bibfnamefont {A.}~\bibnamefont
  {Das}}, \bibinfo {author} {\bibfnamefont {J.}~\bibnamefont {Sabbatini}},\
  and\ \bibinfo {author} {\bibfnamefont {W.~H.}\ \bibnamefont {Zurek}},\
  }\bibfield  {title} {\bibinfo {title} {{Winding up superfluid in a torus via
  Bose Einstein condensation}},\ }\href
  {https://doi.org/https://doi.org/10.1038/srep00352} {\bibfield  {journal}
  {\bibinfo  {journal} {Sci. Rep.}\ }\textbf {\bibinfo {volume} {2}},\ \bibinfo
  {pages} {352} (\bibinfo {year} {2012}{\natexlab{a}})}\BibitemShut {NoStop}%
\bibitem [{\citenamefont {Corman}\ \emph {et~al.}(2014)\citenamefont {Corman},
  \citenamefont {Chomaz}, \citenamefont {Bienaim\'e}, \citenamefont
  {Desbuquois}, \citenamefont {Weitenberg}, \citenamefont {Nascimb\`ene},
  \citenamefont {Dalibard},\ and\ \citenamefont {Beugnon}}]{CormanPRL2014}%
  \BibitemOpen
  \bibfield  {author} {\bibinfo {author} {\bibfnamefont {L.}~\bibnamefont
  {Corman}}, \bibinfo {author} {\bibfnamefont {L.}~\bibnamefont {Chomaz}},
  \bibinfo {author} {\bibfnamefont {T.}~\bibnamefont {Bienaim\'e}}, \bibinfo
  {author} {\bibfnamefont {R.}~\bibnamefont {Desbuquois}}, \bibinfo {author}
  {\bibfnamefont {C.}~\bibnamefont {Weitenberg}}, \bibinfo {author}
  {\bibfnamefont {S.}~\bibnamefont {Nascimb\`ene}}, \bibinfo {author}
  {\bibfnamefont {J.}~\bibnamefont {Dalibard}},\ and\ \bibinfo {author}
  {\bibfnamefont {J.}~\bibnamefont {Beugnon}},\ }\bibfield  {title} {\bibinfo
  {title} {{Quench-Induced Supercurrents in an Annular Bose Gas}},\ }\href
  {https://doi.org/10.1103/PhysRevLett.113.135302} {\bibfield  {journal}
  {\bibinfo  {journal} {Phys. Rev. Lett.}\ }\textbf {\bibinfo {volume} {113}},\
  \bibinfo {pages} {135302} (\bibinfo {year} {2014})}\BibitemShut {NoStop}%
\bibitem [{\citenamefont {Yatsuta}\ \emph {et~al.}(2020)\citenamefont
  {Yatsuta}, \citenamefont {Malomed},\ and\ \citenamefont
  {Yakimenko}}]{YatsutaPRR2020}%
  \BibitemOpen
  \bibfield  {author} {\bibinfo {author} {\bibfnamefont {I.}~\bibnamefont
  {Yatsuta}}, \bibinfo {author} {\bibfnamefont {B.}~\bibnamefont {Malomed}},\
  and\ \bibinfo {author} {\bibfnamefont {A.}~\bibnamefont {Yakimenko}},\
  }\bibfield  {title} {\bibinfo {title} {{Acoustic analog of Hawking radiation
  in quantized circular superflows of Bose-Einstein condensates}},\ }\href
  {https://doi.org/10.1103/PhysRevResearch.2.043065} {\bibfield  {journal}
  {\bibinfo  {journal} {Phys. Rev. Res.}\ }\textbf {\bibinfo {volume} {2}},\
  \bibinfo {pages} {043065} (\bibinfo {year} {2020})}\BibitemShut {NoStop}%
\bibitem [{\citenamefont {Eckel}\ \emph {et~al.}(2018)\citenamefont {Eckel},
  \citenamefont {Kumar}, \citenamefont {Jacobson}, \citenamefont {Spielman},\
  and\ \citenamefont {Campbell}}]{EckelPRX2018}%
  \BibitemOpen
  \bibfield  {author} {\bibinfo {author} {\bibfnamefont {S.}~\bibnamefont
  {Eckel}}, \bibinfo {author} {\bibfnamefont {A.}~\bibnamefont {Kumar}},
  \bibinfo {author} {\bibfnamefont {T.}~\bibnamefont {Jacobson}}, \bibinfo
  {author} {\bibfnamefont {I.~B.}\ \bibnamefont {Spielman}},\ and\ \bibinfo
  {author} {\bibfnamefont {G.~K.}\ \bibnamefont {Campbell}},\ }\bibfield
  {title} {\bibinfo {title} {{A Rapidly Expanding Bose-Einstein Condensate: An
  Expanding Universe in the Lab}},\ }\href
  {https://doi.org/10.1103/PhysRevX.8.021021} {\bibfield  {journal} {\bibinfo
  {journal} {Phys. Rev. X}\ }\textbf {\bibinfo {volume} {8}},\ \bibinfo {pages}
  {021021} (\bibinfo {year} {2018})}\BibitemShut {NoStop}%
\bibitem [{\citenamefont {\"Ohberg}\ and\ \citenamefont
  {Wright}(2019)}]{OhbergPRL2019}%
  \BibitemOpen
  \bibfield  {author} {\bibinfo {author} {\bibfnamefont {P.}~\bibnamefont
  {\"Ohberg}}\ and\ \bibinfo {author} {\bibfnamefont {E.~M.}\ \bibnamefont
  {Wright}},\ }\bibfield  {title} {\bibinfo {title} {{Quantum Time Crystals and
  Interacting Gauge Theories in Atomic Bose-Einstein Condensates}},\ }\href
  {https://doi.org/10.1103/PhysRevLett.123.250402} {\bibfield  {journal}
  {\bibinfo  {journal} {Phys. Rev. Lett.}\ }\textbf {\bibinfo {volume} {123}},\
  \bibinfo {pages} {250402} (\bibinfo {year} {2019})}\BibitemShut {NoStop}%
\bibitem [{\citenamefont {Gupta}\ \emph {et~al.}(2005)\citenamefont {Gupta},
  \citenamefont {Murch}, \citenamefont {Moore}, \citenamefont {Purdy},\ and\
  \citenamefont {Stamper-Kurn}}]{GuptaPRL2005}%
  \BibitemOpen
  \bibfield  {author} {\bibinfo {author} {\bibfnamefont {S.}~\bibnamefont
  {Gupta}}, \bibinfo {author} {\bibfnamefont {K.~W.}\ \bibnamefont {Murch}},
  \bibinfo {author} {\bibfnamefont {K.~L.}\ \bibnamefont {Moore}}, \bibinfo
  {author} {\bibfnamefont {T.~P.}\ \bibnamefont {Purdy}},\ and\ \bibinfo
  {author} {\bibfnamefont {D.~M.}\ \bibnamefont {Stamper-Kurn}},\ }\bibfield
  {title} {\bibinfo {title} {{Bose-Einstein Condensation in a Circular
  Waveguide}},\ }\href {https://doi.org/10.1103/PhysRevLett.95.143201}
  {\bibfield  {journal} {\bibinfo  {journal} {Phys. Rev. Lett.}\ }\textbf
  {\bibinfo {volume} {95}},\ \bibinfo {pages} {143201} (\bibinfo {year}
  {2005})}\BibitemShut {NoStop}%
\bibitem [{\citenamefont {Marti}\ \emph {et~al.}(2015)\citenamefont {Marti},
  \citenamefont {Olf},\ and\ \citenamefont {Stamper-Kurn}}]{MartiPRA2015}%
  \BibitemOpen
  \bibfield  {author} {\bibinfo {author} {\bibfnamefont {G.~E.}\ \bibnamefont
  {Marti}}, \bibinfo {author} {\bibfnamefont {R.}~\bibnamefont {Olf}},\ and\
  \bibinfo {author} {\bibfnamefont {D.~M.}\ \bibnamefont {Stamper-Kurn}},\
  }\bibfield  {title} {\bibinfo {title} {{Collective excitation interferometry
  with a toroidal Bose-Einstein condensate}},\ }\href
  {https://doi.org/10.1103/PhysRevA.91.013602} {\bibfield  {journal} {\bibinfo
  {journal} {Phys. Rev. A}\ }\textbf {\bibinfo {volume} {91}},\ \bibinfo
  {pages} {013602} (\bibinfo {year} {2015})}\BibitemShut {NoStop}%
\bibitem [{\citenamefont {Cooper}\ \emph {et~al.}(2010)\citenamefont {Cooper},
  \citenamefont {Hallwood},\ and\ \citenamefont {Dunningham}}]{CooperPRA2010}%
  \BibitemOpen
  \bibfield  {author} {\bibinfo {author} {\bibfnamefont {J.~J.}\ \bibnamefont
  {Cooper}}, \bibinfo {author} {\bibfnamefont {D.~W.}\ \bibnamefont
  {Hallwood}},\ and\ \bibinfo {author} {\bibfnamefont {J.~A.}\ \bibnamefont
  {Dunningham}},\ }\bibfield  {title} {\bibinfo {title} {Entanglement-enhanced
  atomic gyroscope},\ }\href {https://doi.org/10.1103/PhysRevA.81.043624}
  {\bibfield  {journal} {\bibinfo  {journal} {Phys. Rev. A}\ }\textbf {\bibinfo
  {volume} {81}},\ \bibinfo {pages} {043624} (\bibinfo {year}
  {2010})}\BibitemShut {NoStop}%
\bibitem [{\citenamefont {Pelegri}\ \emph {et~al.}(2018)\citenamefont
  {Pelegri}, \citenamefont {Mompart},\ and\ \citenamefont
  {Ahufinger}}]{PelegriNJP2018}%
  \BibitemOpen
  \bibfield  {author} {\bibinfo {author} {\bibfnamefont {G.}~\bibnamefont
  {Pelegri}}, \bibinfo {author} {\bibfnamefont {J.}~\bibnamefont {Mompart}},\
  and\ \bibinfo {author} {\bibfnamefont {V.}~\bibnamefont {Ahufinger}},\
  }\bibfield  {title} {\bibinfo {title} {{Quantum sensing using imbalanced
  counter-rotating Bose–Einstein condensate modes}},\ }\href
  {https://doi.org/10.1088/1367-2630/aae107} {\bibfield  {journal} {\bibinfo
  {journal} {New J. Phys.}\ }\textbf {\bibinfo {volume} {20}},\ \bibinfo
  {pages} {103001} (\bibinfo {year} {2018})}\BibitemShut {NoStop}%
\bibitem [{\citenamefont {Ramanathan}\ \emph {et~al.}(2011)\citenamefont
  {Ramanathan}, \citenamefont {Wright}, \citenamefont {Muniz}, \citenamefont
  {Zelan}, \citenamefont {Hill}, \citenamefont {Lobb}, \citenamefont
  {Helmerson}, \citenamefont {Phillips},\ and\ \citenamefont
  {Campbell}}]{RamanathanPRL2011}%
  \BibitemOpen
  \bibfield  {author} {\bibinfo {author} {\bibfnamefont {A.}~\bibnamefont
  {Ramanathan}}, \bibinfo {author} {\bibfnamefont {K.~C.}\ \bibnamefont
  {Wright}}, \bibinfo {author} {\bibfnamefont {S.~R.}\ \bibnamefont {Muniz}},
  \bibinfo {author} {\bibfnamefont {M.}~\bibnamefont {Zelan}}, \bibinfo
  {author} {\bibfnamefont {W.~T.}\ \bibnamefont {Hill}}, \bibinfo {author}
  {\bibfnamefont {C.~J.}\ \bibnamefont {Lobb}}, \bibinfo {author}
  {\bibfnamefont {K.}~\bibnamefont {Helmerson}}, \bibinfo {author}
  {\bibfnamefont {W.~D.}\ \bibnamefont {Phillips}},\ and\ \bibinfo {author}
  {\bibfnamefont {G.~K.}\ \bibnamefont {Campbell}},\ }\bibfield  {title}
  {\bibinfo {title} {{Superflow in a Toroidal Bose-Einstein Condensate: An Atom
  Circuit with a Tunable Weak Link}},\ }\href
  {https://doi.org/10.1103/PhysRevLett.106.130401} {\bibfield  {journal}
  {\bibinfo  {journal} {Phys. Rev. Lett.}\ }\textbf {\bibinfo {volume} {106}},\
  \bibinfo {pages} {130401} (\bibinfo {year} {2011})}\BibitemShut {NoStop}%
\bibitem [{\citenamefont {Ryu}\ \emph {et~al.}(2013)\citenamefont {Ryu},
  \citenamefont {Blackburn}, \citenamefont {Blinova},\ and\ \citenamefont
  {Boshier}}]{RyuPRL2013}%
  \BibitemOpen
  \bibfield  {author} {\bibinfo {author} {\bibfnamefont {C.}~\bibnamefont
  {Ryu}}, \bibinfo {author} {\bibfnamefont {P.~W.}\ \bibnamefont {Blackburn}},
  \bibinfo {author} {\bibfnamefont {A.~A.}\ \bibnamefont {Blinova}},\ and\
  \bibinfo {author} {\bibfnamefont {M.~G.}\ \bibnamefont {Boshier}},\
  }\bibfield  {title} {\bibinfo {title} {{Experimental Realization of Josephson
  Junctions for an Atom SQUID}},\ }\href
  {https://doi.org/10.1103/PhysRevLett.111.205301} {\bibfield  {journal}
  {\bibinfo  {journal} {Phys. Rev. Lett.}\ }\textbf {\bibinfo {volume} {111}},\
  \bibinfo {pages} {205301} (\bibinfo {year} {2013})}\BibitemShut {NoStop}%
\bibitem [{\citenamefont {Pandey}\ \emph {et~al.}(2021)\citenamefont {Pandey},
  \citenamefont {Mas}, \citenamefont {Vasilakis},\ and\ \citenamefont {von
  Klitzing}}]{PandeyPRL2021}%
  \BibitemOpen
  \bibfield  {author} {\bibinfo {author} {\bibfnamefont {S.}~\bibnamefont
  {Pandey}}, \bibinfo {author} {\bibfnamefont {H.}~\bibnamefont {Mas}},
  \bibinfo {author} {\bibfnamefont {G.}~\bibnamefont {Vasilakis}},\ and\
  \bibinfo {author} {\bibfnamefont {W.}~\bibnamefont {von Klitzing}},\
  }\bibfield  {title} {\bibinfo {title} {{Atomtronic Matter-Wave Lensing}},\
  }\href {https://doi.org/10.1103/PhysRevLett.126.170402} {\bibfield  {journal}
  {\bibinfo  {journal} {Phys. Rev. Lett.}\ }\textbf {\bibinfo {volume} {126}},\
  \bibinfo {pages} {170402} (\bibinfo {year} {2021})}\BibitemShut {NoStop}%
\bibitem [{\citenamefont {Amico}\ \emph {et~al.}(2022)\citenamefont {Amico},
  \citenamefont {Anderson}, \citenamefont {Boshier}, \citenamefont {Brantut},
  \citenamefont {Kwek}, \citenamefont {Minguzzi},\ and\ \citenamefont {von
  Klitzing}}]{AmicoRMP2022}%
  \BibitemOpen
  \bibfield  {author} {\bibinfo {author} {\bibfnamefont {L.}~\bibnamefont
  {Amico}}, \bibinfo {author} {\bibfnamefont {D.}~\bibnamefont {Anderson}},
  \bibinfo {author} {\bibfnamefont {M.}~\bibnamefont {Boshier}}, \bibinfo
  {author} {\bibfnamefont {J.-P.}\ \bibnamefont {Brantut}}, \bibinfo {author}
  {\bibfnamefont {L.-C.}\ \bibnamefont {Kwek}}, \bibinfo {author}
  {\bibfnamefont {A.}~\bibnamefont {Minguzzi}},\ and\ \bibinfo {author}
  {\bibfnamefont {W.}~\bibnamefont {von Klitzing}},\ }\bibfield  {title}
  {\bibinfo {title} {{Colloquium: Atomtronic circuits: From many-body physics
  to quantum technologies}},\ }\href
  {https://doi.org/10.1103/RevModPhys.94.041001} {\bibfield  {journal}
  {\bibinfo  {journal} {Rev. Mod. Phys.}\ }\textbf {\bibinfo {volume} {94}},\
  \bibinfo {pages} {041001} (\bibinfo {year} {2022})}\BibitemShut {NoStop}%
\bibitem [{\citenamefont {Aghamalyan}\ \emph {et~al.}(2015)\citenamefont
  {Aghamalyan}, \citenamefont {Cominotti}, \citenamefont {Rizzi}, \citenamefont
  {Rossini}, \citenamefont {Hekking}, \citenamefont {Minguzzi}, \citenamefont
  {Kwek},\ and\ \citenamefont {Amico}}]{AghmalyanNJP2015}%
  \BibitemOpen
  \bibfield  {author} {\bibinfo {author} {\bibfnamefont {D.}~\bibnamefont
  {Aghamalyan}}, \bibinfo {author} {\bibfnamefont {M.}~\bibnamefont
  {Cominotti}}, \bibinfo {author} {\bibfnamefont {M.}~\bibnamefont {Rizzi}},
  \bibinfo {author} {\bibfnamefont {D.}~\bibnamefont {Rossini}}, \bibinfo
  {author} {\bibfnamefont {F.}~\bibnamefont {Hekking}}, \bibinfo {author}
  {\bibfnamefont {A.}~\bibnamefont {Minguzzi}}, \bibinfo {author}
  {\bibfnamefont {L.-C.}\ \bibnamefont {Kwek}},\ and\ \bibinfo {author}
  {\bibfnamefont {L.}~\bibnamefont {Amico}},\ }\bibfield  {title} {\bibinfo
  {title} {Coherent superposition of current flows in an atomtronic quantum
  interference device},\ }\href
  {https://doi.org/doi:10.1088/1367-2630/17/4/045023} {\bibfield  {journal}
  {\bibinfo  {journal} {New J. Phys.}\ }\textbf {\bibinfo {volume} {17}},\
  \bibinfo {pages} {045023} (\bibinfo {year} {2015})}\BibitemShut {NoStop}%
\bibitem [{\citenamefont {Ragole}\ and\ \citenamefont
  {Taylor}(2016)}]{RagolePRL2016}%
  \BibitemOpen
  \bibfield  {author} {\bibinfo {author} {\bibfnamefont {S.}~\bibnamefont
  {Ragole}}\ and\ \bibinfo {author} {\bibfnamefont {J.~M.}\ \bibnamefont
  {Taylor}},\ }\bibfield  {title} {\bibinfo {title} {{Interacting Atomic
  Interferometry for Rotation Sensing Approaching the Heisenberg Limit}},\
  }\href {https://doi.org/10.1103/PhysRevLett.117.203002} {\bibfield  {journal}
  {\bibinfo  {journal} {Phys. Rev. Lett.}\ }\textbf {\bibinfo {volume} {117}},\
  \bibinfo {pages} {203002} (\bibinfo {year} {2016})}\BibitemShut {NoStop}%
\bibitem [{\citenamefont {Murray}\ \emph {et~al.}(2013)\citenamefont {Murray},
  \citenamefont {Krygier}, \citenamefont {Edwards}, \citenamefont {Wright},
  \citenamefont {Campbell},\ and\ \citenamefont {Clark}}]{MurrayPRA2013}%
  \BibitemOpen
  \bibfield  {author} {\bibinfo {author} {\bibfnamefont {N.}~\bibnamefont
  {Murray}}, \bibinfo {author} {\bibfnamefont {M.}~\bibnamefont {Krygier}},
  \bibinfo {author} {\bibfnamefont {M.}~\bibnamefont {Edwards}}, \bibinfo
  {author} {\bibfnamefont {K.~C.}\ \bibnamefont {Wright}}, \bibinfo {author}
  {\bibfnamefont {G.~K.}\ \bibnamefont {Campbell}},\ and\ \bibinfo {author}
  {\bibfnamefont {C.~W.}\ \bibnamefont {Clark}},\ }\bibfield  {title} {\bibinfo
  {title} {Probing the circulation of ring-shaped bose-einstein condensates},\
  }\href {https://doi.org/10.1103/PhysRevA.88.053615} {\bibfield  {journal}
  {\bibinfo  {journal} {Phys. Rev. A}\ }\textbf {\bibinfo {volume} {88}},\
  \bibinfo {pages} {053615} (\bibinfo {year} {2013})}\BibitemShut {NoStop}%
\bibitem [{\citenamefont {Kumar}\ \emph {et~al.}(2016)\citenamefont {Kumar},
  \citenamefont {Anderson}, \citenamefont {Phillips}, \citenamefont {Eckel},
  \citenamefont {Campbell},\ and\ \citenamefont {Stringari}}]{KumarNJP2016}%
  \BibitemOpen
  \bibfield  {author} {\bibinfo {author} {\bibfnamefont {A.}~\bibnamefont
  {Kumar}}, \bibinfo {author} {\bibfnamefont {N.}~\bibnamefont {Anderson}},
  \bibinfo {author} {\bibfnamefont {W.~D.}\ \bibnamefont {Phillips}}, \bibinfo
  {author} {\bibfnamefont {S.}~\bibnamefont {Eckel}}, \bibinfo {author}
  {\bibfnamefont {G.~K.}\ \bibnamefont {Campbell}},\ and\ \bibinfo {author}
  {\bibfnamefont {S.}~\bibnamefont {Stringari}},\ }\bibfield  {title} {\bibinfo
  {title} {{Minimally destructive, Doppler measurement of a quantized flow in a
  ring-shaped Bose–Einstein condensate}},\ }\href
  {https://doi.org/10.1088/1367-2630/18/2/025001} {\bibfield  {journal}
  {\bibinfo  {journal} {New J. Phys.}\ }\textbf {\bibinfo {volume} {18}},\
  \bibinfo {pages} {025001} (\bibinfo {year} {2016})}\BibitemShut {NoStop}%
\bibitem [{\citenamefont {Brennecke}\ \emph {et~al.}(2008)\citenamefont
  {Brennecke}, \citenamefont {Ritter}, \citenamefont {Donner},\ and\
  \citenamefont {Esslinger}}]{BrenneckeScience2008}%
  \BibitemOpen
  \bibfield  {author} {\bibinfo {author} {\bibfnamefont {F.}~\bibnamefont
  {Brennecke}}, \bibinfo {author} {\bibfnamefont {S.}~\bibnamefont {Ritter}},
  \bibinfo {author} {\bibfnamefont {T.}~\bibnamefont {Donner}},\ and\ \bibinfo
  {author} {\bibfnamefont {T.}~\bibnamefont {Esslinger}},\ }\bibfield  {title}
  {\bibinfo {title} {{Cavity Optomechanics with a Bose-Einstein Condensate}},\
  }\href {https://doi.org/10.1126/science.116321} {\bibfield  {journal}
  {\bibinfo  {journal} {Science}\ }\textbf {\bibinfo {volume} {322}},\ \bibinfo
  {pages} {235} (\bibinfo {year} {2008})}\BibitemShut {NoStop}%
\bibitem [{\citenamefont {Aspelmeyer}\ \emph {et~al.}(2014)\citenamefont
  {Aspelmeyer}, \citenamefont {Kippenberg},\ and\ \citenamefont
  {Marquardt}}]{AspelmeyerRMP2014}%
  \BibitemOpen
  \bibfield  {author} {\bibinfo {author} {\bibfnamefont {M.}~\bibnamefont
  {Aspelmeyer}}, \bibinfo {author} {\bibfnamefont {T.~J.}\ \bibnamefont
  {Kippenberg}},\ and\ \bibinfo {author} {\bibfnamefont {F.}~\bibnamefont
  {Marquardt}},\ }\bibfield  {title} {\bibinfo {title} {Cavity optomechanics},\
  }\href {https://doi.org/10.1103/RevModPhys.86.1391} {\bibfield  {journal}
  {\bibinfo  {journal} {Rev. Mod. Phys.}\ }\textbf {\bibinfo {volume} {86}},\
  \bibinfo {pages} {1391} (\bibinfo {year} {2014})}\BibitemShut {NoStop}%
\bibitem [{\citenamefont {Abbott}\ \emph {et~al.}(2016)\citenamefont {Abbott}
  \emph {et~al.}}]{AbbottPRL2016}%
  \BibitemOpen
  \bibfield  {author} {\bibinfo {author} {\bibfnamefont {B.~P.}\ \bibnamefont
  {Abbott}} \emph {et~al.} (\bibinfo {collaboration} {LIGO Scientific
  Collaboration and Virgo Collaboration}),\ }\bibfield  {title} {\bibinfo
  {title} {{Observation of Gravitational Waves from a Binary Black Hole
  Merger}},\ }\href {https://doi.org/10.1103/PhysRevLett.116.061102} {\bibfield
   {journal} {\bibinfo  {journal} {Phys. Rev. Lett.}\ }\textbf {\bibinfo
  {volume} {116}},\ \bibinfo {pages} {061102} (\bibinfo {year}
  {2016})}\BibitemShut {NoStop}%
\bibitem [{\citenamefont {He}\ \emph {et~al.}(2009)\citenamefont {He},
  \citenamefont {Xu}, \citenamefont {Wang},\ and\ \citenamefont
  {Zhan}}]{HeOESLMOL}%
  \BibitemOpen
  \bibfield  {author} {\bibinfo {author} {\bibfnamefont {X.}~\bibnamefont
  {He}}, \bibinfo {author} {\bibfnamefont {P.}~\bibnamefont {Xu}}, \bibinfo
  {author} {\bibfnamefont {J.}~\bibnamefont {Wang}},\ and\ \bibinfo {author}
  {\bibfnamefont {M.}~\bibnamefont {Zhan}},\ }\bibfield  {title} {\bibinfo
  {title} {Rotating single atoms in a ring lattice generated by a spatial light
  modulator},\ }\href {https://doi.org/10.1364/OE.17.021007} {\bibfield
  {journal} {\bibinfo  {journal} {Opt. Express}\ }\textbf {\bibinfo {volume}
  {17}},\ \bibinfo {pages} {21007} (\bibinfo {year} {2009})}\BibitemShut
  {NoStop}%
\bibitem [{\citenamefont {Brand}\ and\ \citenamefont
  {Reinhardt}(2001)}]{BrandJPB2001}%
  \BibitemOpen
  \bibfield  {author} {\bibinfo {author} {\bibfnamefont {J.}~\bibnamefont
  {Brand}}\ and\ \bibinfo {author} {\bibfnamefont {W.~P.}\ \bibnamefont
  {Reinhardt}},\ }\bibfield  {title} {\bibinfo {title} {{Generating ring
  currents, solitons and vortices by stirring a Bose–Einstein condensate in a
  toroidal trap}},\ }\href {https://doi.org/10.1088/0953-4075/34/4/105}
  {\bibfield  {journal} {\bibinfo  {journal} {J. Phys. B: At. Mol. Opt. Phys.}\
  }\textbf {\bibinfo {volume} {34}},\ \bibinfo {pages} {L113} (\bibinfo {year}
  {2001})}\BibitemShut {NoStop}%
\bibitem [{\citenamefont {Kavoulakis}(2003)}]{KavoulakisPRA2003}%
  \BibitemOpen
  \bibfield  {author} {\bibinfo {author} {\bibfnamefont {G.~M.}\ \bibnamefont
  {Kavoulakis}},\ }\bibfield  {title} {\bibinfo {title} {{Bose-Einstein
  condensates with attractive interactions on a ring}},\ }\href
  {https://doi.org/10.1103/PhysRevA.67.011601} {\bibfield  {journal} {\bibinfo
  {journal} {Phys. Rev. A}\ }\textbf {\bibinfo {volume} {67}},\ \bibinfo
  {pages} {011601} (\bibinfo {year} {2003})}\BibitemShut {NoStop}%
\bibitem [{\citenamefont {Toikka}\ \emph {et~al.}(2013)\citenamefont {Toikka},
  \citenamefont {Kärki},\ and\ \citenamefont {Suominen}}]{ToikkaJPB2013}%
  \BibitemOpen
  \bibfield  {author} {\bibinfo {author} {\bibfnamefont {L.~A.}\ \bibnamefont
  {Toikka}}, \bibinfo {author} {\bibfnamefont {O.}~\bibnamefont {Kärki}},\
  and\ \bibinfo {author} {\bibfnamefont {K.-A.}\ \bibnamefont {Suominen}},\
  }\bibfield  {title} {\bibinfo {title} {{Creation and revival of ring dark
  solitons in an annular Bose–Einstein condensate}},\ }\href
  {https://doi.org/10.1088/0953-4075/47/2/021002} {\bibfield  {journal}
  {\bibinfo  {journal} {Journal of Physics B: Atomic, Molecular and Optical
  Physics}\ }\textbf {\bibinfo {volume} {47}},\ \bibinfo {pages} {021002}
  (\bibinfo {year} {2013})}\BibitemShut {NoStop}%
\bibitem [{\citenamefont {McDonald}\ \emph {et~al.}(2014)\citenamefont
  {McDonald}, \citenamefont {Kuhn}, \citenamefont {Hardman}, \citenamefont
  {Bennetts}, \citenamefont {Everitt}, \citenamefont {Altin}, \citenamefont
  {Debs}, \citenamefont {Close},\ and\ \citenamefont
  {Robins}}]{McDonaldPRL2014}%
  \BibitemOpen
  \bibfield  {author} {\bibinfo {author} {\bibfnamefont {G.~D.}\ \bibnamefont
  {McDonald}}, \bibinfo {author} {\bibfnamefont {C.~C.~N.}\ \bibnamefont
  {Kuhn}}, \bibinfo {author} {\bibfnamefont {K.~S.}\ \bibnamefont {Hardman}},
  \bibinfo {author} {\bibfnamefont {S.}~\bibnamefont {Bennetts}}, \bibinfo
  {author} {\bibfnamefont {P.~J.}\ \bibnamefont {Everitt}}, \bibinfo {author}
  {\bibfnamefont {P.~A.}\ \bibnamefont {Altin}}, \bibinfo {author}
  {\bibfnamefont {J.~E.}\ \bibnamefont {Debs}}, \bibinfo {author}
  {\bibfnamefont {J.~D.}\ \bibnamefont {Close}},\ and\ \bibinfo {author}
  {\bibfnamefont {N.~P.}\ \bibnamefont {Robins}},\ }\bibfield  {title}
  {\bibinfo {title} {{Bright Solitonic Matter-Wave Interferometer}},\ }\href
  {https://doi.org/10.1103/PhysRevLett.113.013002} {\bibfield  {journal}
  {\bibinfo  {journal} {Phys. Rev. Lett.}\ }\textbf {\bibinfo {volume} {113}},\
  \bibinfo {pages} {013002} (\bibinfo {year} {2014})}\BibitemShut {NoStop}%
\bibitem [{\citenamefont {Helm}\ \emph {et~al.}(2015)\citenamefont {Helm},
  \citenamefont {Cornish},\ and\ \citenamefont {Gardiner}}]{HelmPRL2015}%
  \BibitemOpen
  \bibfield  {author} {\bibinfo {author} {\bibfnamefont {J.~L.}\ \bibnamefont
  {Helm}}, \bibinfo {author} {\bibfnamefont {S.~L.}\ \bibnamefont {Cornish}},\
  and\ \bibinfo {author} {\bibfnamefont {S.~A.}\ \bibnamefont {Gardiner}},\
  }\bibfield  {title} {\bibinfo {title} {{Sagnac Interferometry Using Bright
  Matter-Wave Solitons}},\ }\href
  {https://doi.org/10.1103/PhysRevLett.114.134101} {\bibfield  {journal}
  {\bibinfo  {journal} {Phys. Rev. Lett.}\ }\textbf {\bibinfo {volume} {114}},\
  \bibinfo {pages} {134101} (\bibinfo {year} {2015})}\BibitemShut {NoStop}%
\bibitem [{\citenamefont {Gallucci}\ and\ \citenamefont
  {Proukakis}(2016)}]{GalluciNJP2016}%
  \BibitemOpen
  \bibfield  {author} {\bibinfo {author} {\bibfnamefont {D.}~\bibnamefont
  {Gallucci}}\ and\ \bibinfo {author} {\bibfnamefont {N.~P.}\ \bibnamefont
  {Proukakis}},\ }\bibfield  {title} {\bibinfo {title} {{Engineering dark
  solitary waves in ring-trap Bose–Einstein condensates}},\ }\href
  {https://doi.org/10.1088/1367-2630/18/2/025004} {\bibfield  {journal}
  {\bibinfo  {journal} {New J. Phys.}\ }\textbf {\bibinfo {volume} {18}},\
  \bibinfo {pages} {025004} (\bibinfo {year} {2016})}\BibitemShut {NoStop}%
\bibitem [{\citenamefont {Jezek}\ \emph {et~al.}(2016)\citenamefont {Jezek},
  \citenamefont {Capuzzi},\ and\ \citenamefont {Cataldo}}]{JezekPRA2016}%
  \BibitemOpen
  \bibfield  {author} {\bibinfo {author} {\bibfnamefont {D.~M.}\ \bibnamefont
  {Jezek}}, \bibinfo {author} {\bibfnamefont {P.}~\bibnamefont {Capuzzi}},\
  and\ \bibinfo {author} {\bibfnamefont {H.~M.}\ \bibnamefont {Cataldo}},\
  }\bibfield  {title} {\bibinfo {title} {{Dark-soliton collisions in a toroidal
  Bose-Einstein condensate}},\ }\href
  {https://doi.org/10.1103/PhysRevA.93.023601} {\bibfield  {journal} {\bibinfo
  {journal} {Phys. Rev. A}\ }\textbf {\bibinfo {volume} {93}},\ \bibinfo
  {pages} {023601} (\bibinfo {year} {2016})}\BibitemShut {NoStop}%
\bibitem [{\citenamefont {Cataldo}\ and\ \citenamefont
  {Jezek}(2016)}]{CataldoEPJD2016}%
  \BibitemOpen
  \bibfield  {author} {\bibinfo {author} {\bibfnamefont {H.}~\bibnamefont
  {Cataldo}}\ and\ \bibinfo {author} {\bibfnamefont {D.~M.}\ \bibnamefont
  {Jezek}},\ }\bibfield  {title} {\bibinfo {title} {{Collisional dynamics of
  multiple dark solitons in a toroidal Bose-Einstein condensate: Quasiparticle
  picture}},\ }\href {https://doi.org/10.1140/epjd/e2018-90222-8} {\bibfield
  {journal} {\bibinfo  {journal} {Euro. Phys. J. D}\ }\textbf {\bibinfo
  {volume} {72}},\ \bibinfo {pages} {213} (\bibinfo {year} {2016})}\BibitemShut
  {NoStop}%
\bibitem [{\citenamefont {Safaei}\ \emph {et~al.}(2019)\citenamefont {Safaei},
  \citenamefont {Kwek}, \citenamefont {Dumke},\ and\ \citenamefont
  {Amico}}]{SafaeiPRA2019}%
  \BibitemOpen
  \bibfield  {author} {\bibinfo {author} {\bibfnamefont {S.}~\bibnamefont
  {Safaei}}, \bibinfo {author} {\bibfnamefont {L.-C.}\ \bibnamefont {Kwek}},
  \bibinfo {author} {\bibfnamefont {R.}~\bibnamefont {Dumke}},\ and\ \bibinfo
  {author} {\bibfnamefont {L.}~\bibnamefont {Amico}},\ }\bibfield  {title}
  {\bibinfo {title} {Monitoring currents in cold-atom circuits},\ }\href
  {https://doi.org/10.1103/PhysRevA.100.013621} {\bibfield  {journal} {\bibinfo
   {journal} {Phys. Rev. A}\ }\textbf {\bibinfo {volume} {100}},\ \bibinfo
  {pages} {013621} (\bibinfo {year} {2019})}\BibitemShut {NoStop}%
\bibitem [{\citenamefont {Lim}\ \emph {et~al.}(2022)\citenamefont {Lim},
  \citenamefont {Lee}, \citenamefont {Goo}, \citenamefont {Bae},\ and\
  \citenamefont {Shin}}]{Lim2022Shedding}%
  \BibitemOpen
  \bibfield  {author} {\bibinfo {author} {\bibfnamefont {Y.}~\bibnamefont
  {Lim}}, \bibinfo {author} {\bibfnamefont {Y.}~\bibnamefont {Lee}}, \bibinfo
  {author} {\bibfnamefont {J.}~\bibnamefont {Goo}}, \bibinfo {author}
  {\bibfnamefont {D.}~\bibnamefont {Bae}},\ and\ \bibinfo {author}
  {\bibfnamefont {Y.}~\bibnamefont {Shin}},\ }\bibfield  {title} {\bibinfo
  {title} {{Vortex shedding frequency of a moving obstacle in a Bose-Einstein
  condensate}},\ }\href@noop {} {\bibfield  {journal} {\bibinfo  {journal} {New
  Journal of Physics}\ }\textbf {\bibinfo {volume} {24}},\ \bibinfo {pages}
  {083020} (\bibinfo {year} {2022})}\BibitemShut {NoStop}%
\bibitem [{\citenamefont {Stoof}(1999)}]{stoof1999coherent}%
  \BibitemOpen
  \bibfield  {author} {\bibinfo {author} {\bibfnamefont {H.~T.}\ \bibnamefont
  {Stoof}},\ }\bibfield  {title} {\bibinfo {title} {Coherent versus incoherent
  dynamics during bose-einstein condensation in atomic gases},\ }\href@noop {}
  {\bibfield  {journal} {\bibinfo  {journal} {Journal of low temperature
  physics}\ }\textbf {\bibinfo {volume} {114}},\ \bibinfo {pages} {11}
  (\bibinfo {year} {1999})}\BibitemShut {NoStop}%
\bibitem [{\citenamefont {Stoof}\ and\ \citenamefont
  {Bijlsma}(2001)}]{stoof2001dynamics}%
  \BibitemOpen
  \bibfield  {author} {\bibinfo {author} {\bibfnamefont {H.}~\bibnamefont
  {Stoof}}\ and\ \bibinfo {author} {\bibfnamefont {M.}~\bibnamefont
  {Bijlsma}},\ }\bibfield  {title} {\bibinfo {title} {Dynamics of fluctuating
  bose--einstein condensates},\ }\href@noop {} {\bibfield  {journal} {\bibinfo
  {journal} {Journal of low temperature physics}\ }\textbf {\bibinfo {volume}
  {124}},\ \bibinfo {pages} {431} (\bibinfo {year} {2001})}\BibitemShut
  {NoStop}%
\bibitem [{\citenamefont {Kubo}(1966)}]{kubo1966fluctuation}%
  \BibitemOpen
  \bibfield  {author} {\bibinfo {author} {\bibfnamefont {R.}~\bibnamefont
  {Kubo}},\ }\bibfield  {title} {\bibinfo {title} {The fluctuation-dissipation
  theorem},\ }\href@noop {} {\bibfield  {journal} {\bibinfo  {journal} {Rep.
  Prog. Phys.}\ }\textbf {\bibinfo {volume} {29}},\ \bibinfo {pages} {255}
  (\bibinfo {year} {1966})}\BibitemShut {NoStop}%
\bibitem [{\citenamefont {Mithun}\ \emph {et~al.}(2018)\citenamefont {Mithun},
  \citenamefont {Ganguli}, \citenamefont {Raychaudhuri},\ and\ \citenamefont
  {Dey}}]{mithun2018signatures}%
  \BibitemOpen
  \bibfield  {author} {\bibinfo {author} {\bibfnamefont {T.}~\bibnamefont
  {Mithun}}, \bibinfo {author} {\bibfnamefont {S.~C.}\ \bibnamefont {Ganguli}},
  \bibinfo {author} {\bibfnamefont {P.}~\bibnamefont {Raychaudhuri}},\ and\
  \bibinfo {author} {\bibfnamefont {B.}~\bibnamefont {Dey}},\ }\bibfield
  {title} {\bibinfo {title} {{Signatures of two-step impurity-mediated vortex
  lattice melting in Bose-Einstein condensates}},\ }\href@noop {} {\bibfield
  {journal} {\bibinfo  {journal} {Europhysics Letters}\ }\textbf {\bibinfo
  {volume} {123}},\ \bibinfo {pages} {20004} (\bibinfo {year}
  {2018})}\BibitemShut {NoStop}%
\bibitem [{\citenamefont {Das}\ \emph {et~al.}(2012{\natexlab{b}})\citenamefont
  {Das}, \citenamefont {Sabbatini},\ and\ \citenamefont
  {Zurek}}]{das2012winding}%
  \BibitemOpen
  \bibfield  {author} {\bibinfo {author} {\bibfnamefont {A.}~\bibnamefont
  {Das}}, \bibinfo {author} {\bibfnamefont {J.}~\bibnamefont {Sabbatini}},\
  and\ \bibinfo {author} {\bibfnamefont {W.~H.}\ \bibnamefont {Zurek}},\
  }\bibfield  {title} {\bibinfo {title} {{Winding up superfluid in a torus via
  Bose Einstein condensation}},\ }\href@noop {} {\bibfield  {journal} {\bibinfo
   {journal} {Scientific reports}\ }\textbf {\bibinfo {volume} {2}},\ \bibinfo
  {pages} {1} (\bibinfo {year} {2012}{\natexlab{b}})}\BibitemShut {NoStop}%
\bibitem [{\citenamefont {Dutykh}(2016)}]{dutykh2016brief}%
  \BibitemOpen
  \bibfield  {author} {\bibinfo {author} {\bibfnamefont {D.}~\bibnamefont
  {Dutykh}},\ }\bibfield  {title} {\bibinfo {title} {A brief introduction to
  pseudo-spectral methods: application to diffusion problems},\ }\href@noop {}
  {\bibfield  {journal} {\bibinfo  {journal} {arXiv preprint arXiv:1606.05432}\
  } (\bibinfo {year} {2016})}\BibitemShut {NoStop}%
\bibitem [{\citenamefont {Tan}\ and\ \citenamefont
  {Chen}(2012)}]{tan2012general}%
  \BibitemOpen
  \bibfield  {author} {\bibinfo {author} {\bibfnamefont {D.}~\bibnamefont
  {Tan}}\ and\ \bibinfo {author} {\bibfnamefont {Z.}~\bibnamefont {Chen}},\
  }\bibfield  {title} {\bibinfo {title} {On a general formula of fourth order
  runge-kutta method},\ }\href@noop {} {\bibfield  {journal} {\bibinfo
  {journal} {Journal of Mathematical Science \& Mathematics Education}\
  }\textbf {\bibinfo {volume} {7}},\ \bibinfo {pages} {1} (\bibinfo {year}
  {2012})}\BibitemShut {NoStop}%
\bibitem [{\citenamefont {Bao}\ \emph {et~al.}(2003)\citenamefont {Bao},
  \citenamefont {Jaksch},\ and\ \citenamefont {Markowich}}]{bao2003numerical}%
  \BibitemOpen
  \bibfield  {author} {\bibinfo {author} {\bibfnamefont {W.}~\bibnamefont
  {Bao}}, \bibinfo {author} {\bibfnamefont {D.}~\bibnamefont {Jaksch}},\ and\
  \bibinfo {author} {\bibfnamefont {P.~A.}\ \bibnamefont {Markowich}},\
  }\bibfield  {title} {\bibinfo {title} {{Numerical solution of the
  Gross--Pitaevskii equation for Bose-Einstein condensation}},\ }\href@noop {}
  {\bibfield  {journal} {\bibinfo  {journal} {Journal of Computational
  Physics}\ }\textbf {\bibinfo {volume} {187}},\ \bibinfo {pages} {318}
  (\bibinfo {year} {2003})}\BibitemShut {NoStop}%
\bibitem [{\citenamefont {Bhattacharya}(2015)}]{bhattacharya2015rotational}%
  \BibitemOpen
  \bibfield  {author} {\bibinfo {author} {\bibfnamefont {M.}~\bibnamefont
  {Bhattacharya}},\ }\bibfield  {title} {\bibinfo {title} {Rotational cavity
  optomechanics},\ }\href@noop {} {\bibfield  {journal} {\bibinfo  {journal}
  {JOSA B}\ }\textbf {\bibinfo {volume} {32}},\ \bibinfo {pages} {B55}
  (\bibinfo {year} {2015})}\BibitemShut {NoStop}%
\bibitem [{\citenamefont {Al~Khawaja}\ \emph {et~al.}(2002)\citenamefont
  {Al~Khawaja}, \citenamefont {Stoof}, \citenamefont {Hulet}, \citenamefont
  {Strecker},\ and\ \citenamefont {Partridge}}]{Khawaja2002}%
  \BibitemOpen
  \bibfield  {author} {\bibinfo {author} {\bibfnamefont {U.}~\bibnamefont
  {Al~Khawaja}}, \bibinfo {author} {\bibfnamefont {H.~T.~C.}\ \bibnamefont
  {Stoof}}, \bibinfo {author} {\bibfnamefont {R.~G.}\ \bibnamefont {Hulet}},
  \bibinfo {author} {\bibfnamefont {K.~E.}\ \bibnamefont {Strecker}},\ and\
  \bibinfo {author} {\bibfnamefont {G.~B.}\ \bibnamefont {Partridge}},\
  }\bibfield  {title} {\bibinfo {title} {Bright soliton trains of trapped
  bose-einstein condensates},\ }\href
  {https://doi.org/10.1103/PhysRevLett.89.200404} {\bibfield  {journal}
  {\bibinfo  {journal} {Phys. Rev. Lett.}\ }\textbf {\bibinfo {volume} {89}},\
  \bibinfo {pages} {200404} (\bibinfo {year} {2002})}\BibitemShut {NoStop}%
\bibitem [{\citenamefont {Gangwar}\ \emph {et~al.}(2022)\citenamefont
  {Gangwar}, \citenamefont {Ravisankar}, \citenamefont {Muruganandam},\ and\
  \citenamefont {Mishra}}]{Gangwar2022}%
  \BibitemOpen
  \bibfield  {author} {\bibinfo {author} {\bibfnamefont {S.}~\bibnamefont
  {Gangwar}}, \bibinfo {author} {\bibfnamefont {R.}~\bibnamefont {Ravisankar}},
  \bibinfo {author} {\bibfnamefont {P.}~\bibnamefont {Muruganandam}},\ and\
  \bibinfo {author} {\bibfnamefont {P.~K.}\ \bibnamefont {Mishra}},\ }\bibfield
   {title} {\bibinfo {title} {Dynamics of quantum solitons in lee-huang-yang
  spin-orbit-coupled bose-einstein condensates},\ }\href
  {https://doi.org/10.1103/PhysRevA.106.063315} {\bibfield  {journal} {\bibinfo
   {journal} {Phys. Rev. A}\ }\textbf {\bibinfo {volume} {106}},\ \bibinfo
  {pages} {063315} (\bibinfo {year} {2022})}\BibitemShut {NoStop}%
\bibitem [{\citenamefont {Nguyen}\ \emph {et~al.}(2014)\citenamefont {Nguyen},
  \citenamefont {Dyke}, \citenamefont {Luo}, \citenamefont {Malomed},\ and\
  \citenamefont {Hulet}}]{nguyen2014collisions}%
  \BibitemOpen
  \bibfield  {author} {\bibinfo {author} {\bibfnamefont {J.~H.}\ \bibnamefont
  {Nguyen}}, \bibinfo {author} {\bibfnamefont {P.}~\bibnamefont {Dyke}},
  \bibinfo {author} {\bibfnamefont {D.}~\bibnamefont {Luo}}, \bibinfo {author}
  {\bibfnamefont {B.~A.}\ \bibnamefont {Malomed}},\ and\ \bibinfo {author}
  {\bibfnamefont {R.~G.}\ \bibnamefont {Hulet}},\ }\bibfield  {title} {\bibinfo
  {title} {Collisions of matter-wave solitons},\ }\href@noop {} {\bibfield
  {journal} {\bibinfo  {journal} {Nature Physics}\ }\textbf {\bibinfo {volume}
  {10}},\ \bibinfo {pages} {918} (\bibinfo {year} {2014})}\BibitemShut
  {NoStop}%
\end{thebibliography}%



\end{document}